\documentclass[%
 reprint,
superscriptaddress,
amsmath,amssymb,aps,prx,floatfix,multicol]{revtex4-2}

\usepackage{bbm, bm, mathtools, hyperref, graphicx, xcolor} %My packages
\newcommand{\edit}[1]{#1}

\newcommand{\Id}{\mathbbm{1}}
\newcommand{\subs}[2]{{#1}_{\mathrm{#2}}}
\newcommand{\sups}[2]{{#1}^{\mathrm{#2}}}

\newcommand{\op}[1]{\hat{#1}}
\newcommand{\Tr}{\mathrm{Tr}}
\newcommand{\proj}[1]{\left | #1 \rangle\langle #1 \right |}
\newcommand{\ket}[1]{\left | #1 \right \rangle}
\newcommand{\bra}[1]{\left \langle #1 \right |}
\newcommand{\braket}[1]{\left \langle #1 \right \rangle}

\begin{document}

\title{
    Entangled quantum cellular automata, physical complexity, and Goldilocks
    rules}

\author{Logan E. Hillberry}
\affiliation{
    Department of Physics, Colorado School of Mines, Golden, CO 80401, USA
}
\affiliation{
    Center for Nonlinear Dynamics, The University of Texas at Austin, Austin,
    TX 78712, USA
}
\author{Matthew T. Jones}
\affiliation{
    Department of Physics, Colorado School of Mines, Golden, CO 80401, USA
}
\author{David L. Vargas}
\affiliation{
    Department of Physics, Colorado School of Mines, Golden, CO 80401, USA
}
\author{Patrick Rall}
\affiliation{
    Quantum Information Center, The University of Texas at Austin, Austin, TX
    78712, USA
}
\affiliation{
    Institute for Quantum Information and Matter, California Institute of
    Technology, Pasadena, CA 91125, USA
}
\author{Nicole Yunger Halpern}
\affiliation{
    Institute for Quantum Information and Matter, California Institute of
    Technology, Pasadena, CA 91125, USA
}
\affiliation{
    Harvard-Smithsonian Center for Astrophysics, Cambridge, MA 02138, USA
}
\affiliation{
    Department of Physics, Harvard University, Cambridge, MA 02138, USA
}
\affiliation{
    Research Laboratory of Electronics, Massachusetts Institute of
    Technology, Cambridge, MA 02139, USA
}
\affiliation{Center for Theoretical Physics, Massachusetts Institute of Technology, Cambridge, Massachusetts 02139, USA}
% \affiliation{Joint Center for Quantum Information and Computer Science, NIST and University of Maryland, College Park, MD 20742, USA}
\affiliation{Institute for Physical Science and Technology, University of Maryland, College Park, MD 20742, USA}
\author{Ning Bao}
\affiliation{
    Institute for Quantum Information and Matter, California Institute of
    Technology, Pasadena, CA 91125, USA
}
\affiliation{
    Berkeley Center for Theoretical Physics, University of California,
    Berkeley, CA 94720, USA
}
\affiliation{Computational-Science Initiative, Brookhaven National Laboratory, Upton, NY 11973, USA}
\author{Simone Notarnicola}
\affiliation{
    Dipartimento di Fisica e Astronomia, Universit\`a degli Studi di
    Padova, I-35131 Italy
}
\affiliation{
    Istituto Nazionale di Fisica Nucleare (INFN), Sezione di Padova, I-35131
    Italy
}
\author{Simone Montangero}
\affiliation{
    Istituto Nazionale di Fisica Nucleare (INFN), Sezione di Padova, I-35131
    Italy
}
\affiliation{
    Dipartimento di Fisica e Astronomia, Universit\`a degli Studi di
    Padova, I-35131 Italy}
\affiliation{
    Theoretische Physik, Universit\"at des Saarlandes, D-66123 Saarbr\"ucken,
    Germany
}
\author{Lincoln D. Carr}
\affiliation{
    Department of Physics, Colorado School of Mines, Golden, CO 80401, USA
}
\email[]{lcarr@mines.edu}

\date{\today}

\begin{abstract}
Cellular automata are interacting classical bits that display diverse emergent
behaviors, from fractals to random-number generators to Turing-complete
computation. We discover that quantum cellular automata (QCA) can exhibit
complexity in the sense of the complexity science that describes biology,
sociology, and economics. QCA exhibit complexity when evolving under
``Goldilocks rules'' that we define by balancing activity and stasis. Our
Goldilocks rules generate robust dynamical features (entangled breathers),
network structure and dynamics consistent with complexity, and persistent
entropy fluctuations. Present-day experimental platforms---Rydberg arrays,
trapped ions, and superconducting qubits---can implement our Goldilocks
protocols, making testable the link between complexity science and quantum
computation exposed by our QCA. \\ MIT-CTP/5277
\end{abstract}

\maketitle

\section{\label{sec:introduction}
    Introduction: Physical complexity in quantum systems}

\begin{figure}[t]
\begin{centering}
\includegraphics[width=0.45\textwidth]{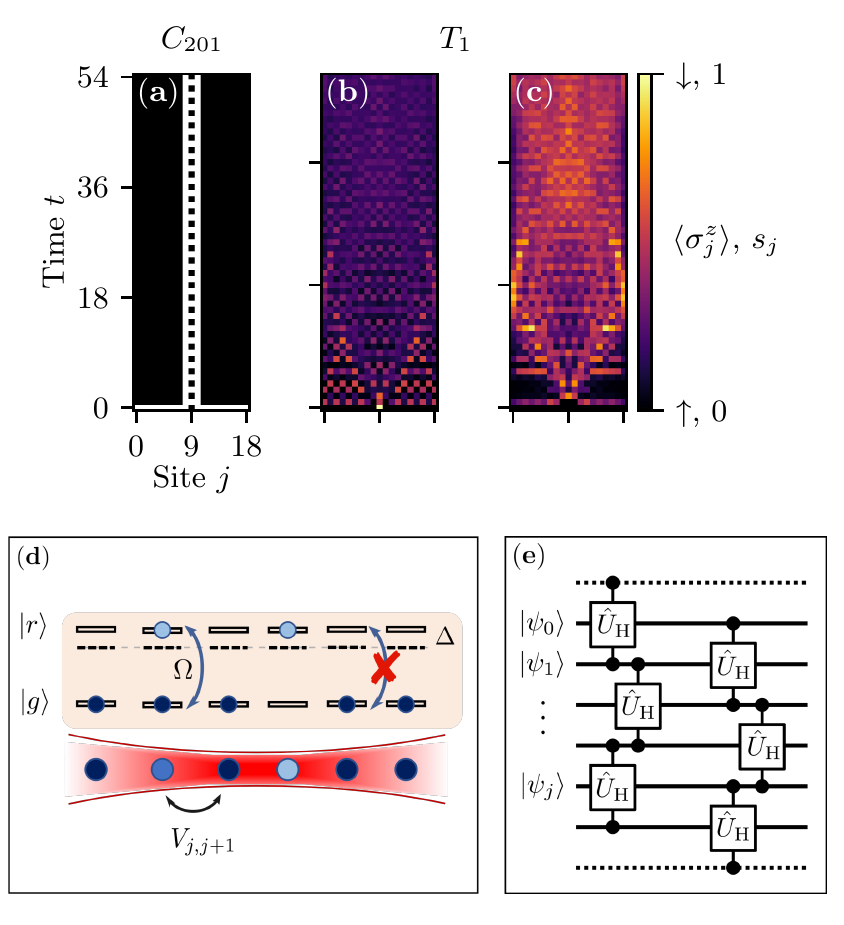}
\caption{
    (a) The classical-elementary-cellular-automaton rule $C_{201}$ flips a
        classical bit (black=1, white=0) if and only if its nearest neighbors
        are 0s~\cite{Wolfram1983Statistical}.  This $L{=}19$-site chain is
        initialized with a 1 centered in 0s.  A periodic pattern, called a
        \emph{blinker}, propagates upward in discrete time steps.
    (b) We extend $C_{201}$, for QCA, to rule $T_1$, defined in the main text.
        The classical bit becomes a quantum bit in an $L{=}19$-site chain.
        The average spin $\braket{\op{\sigma}_j^z}$ has richer discrete-time
        dynamics, oscillating between the classical extremes of 0 (black)
        and 1 (white) and filling the lattice.  A truly quantum analog
        of the classical blinker---an entangled breather---appears in
        Figs.~\ref{fig2}(c)-(d).
    (c) The quantum lattice evolves into a high-entropy entangled state. $s_j$
        denotes the site-$j$ von Neumann entropy.
    (d) An analog quantum computer has implemented rule $T_1$, of which the
        recently named PXP model is a particular case. Optical tweezers were
        used to trap a Rydberg-atom chain~\cite{bernien2017probing}.
    (e) A quantum circuit can realize digital $T_1$ dynamics.  One QCA time
        step requires two layers of controlled-controlled-Hadamard gates.  The
        first layer evolves even-index qubits; and the second layer, odd-index.
        The dashed line represents the boundary qubits, which remain fixed in
        $\ket{0}$'s.
    \label{fig1}
    }
\end{centering}
\end{figure}

Classical cellular automata evolve bit strings (sequences of
1s and 0s) via simple rules that generate diverse emergent
behaviors~\cite{lindgren1988complexity, cook2004universality,
wolfram1985cryptography, Wolfram1983Statistical}.  \emph{Quantum} cellular
automata (QCA) evolve quantum bits (qubits), supporting superpositions and
entanglement (Fig.~\ref{fig1}).  In computer science, complexity characterizes
the number of steps in an algorithm.  The algorithmic-complexity perspective on
cellular automata---quantum~\cite{farrelly2020review,Arrighi2019overview} and
classical~\cite{cook2004universality,sarkar2000brief,di2008computational}---has
received much attention. In contrast, \emph{physical
complexity}~\cite{Bleh2012quantum, hillberry2016entanglement,
vargas2016quantum} characterizes emergent behaviors, including materials'
rigidity, spontaneous symmetry breaking~\cite{anderson1972more,pwanderson1984},
and self-organized criticality~\cite{jensen1998self}. Such emergent
behaviors are exhibited by biological, social, and economic
systems~\cite{Turcotte2002self}.

\edit{
Complexity science is primarily phenomenological and so does not admit of
an axiomatic definition~\cite{standish2008concept}.  Instead, a system is
regarded as complex if data analysis reveals multiple signals widely associated
with complexity~\cite{Newman2003, Csete2002}.  Examples include natural
selection~\cite{adami2002complexity}, diversity~\cite{doebeli2010complexity}
power-law statistics~\cite{jensen1998self}, robustness-fragility
tradeoffs~\cite{carlson2002complexity}, and complex networks~\cite{Newman2003,
Csete2002, reka2002statistical}.  Physical complexity has been observed at the
classical and biological scales.  Does it arise in the quantum dynamical regime?
}

\edit{
We answer affirmatively, showing that QCA can exhibit physical complexity when
subject to evolutions that we term \emph{Goldilocks rules}.  A Goldilocks rule
changes a qubit's state when, and only when, half the qubit's neighbors occupy
the $+1$ $\op{\sigma}_z$ eigenstate, $\ket{1}$. If too many neighbors occupy
$\ket{1}$, or too few neighbors, the central qubit remains static. Goldilocks
rules balance evolution with stasis.  This tradeoff parallels the tradeoffs in
traditional complex systems~\cite{carlson2002complexity}.  Hence one should
expect Goldilocks rules to produce complexity.  They do, we discover---even
in some of the simplest QCA constructable: unitary, one-dimensional (1D)
nearest-neighbor circuits.
}

\edit{
We quantify QCA's complexity using quantum generalizations
of measures used to quantify the brain's complexity in
electroencephalogram/functional-magnetic-resonance-imaging
(EEG/fMRI)~\cite{Bullmore2009, jin2012reorganization}: mutual-information
networks.  For each pair $(j, k)$ of qubits, we calculate the mutual
information $M_{jk}$ between them.  We regard the qubits as a graph's nodes
and the $M_{jk}$ as the links' weights.  The nodes and links form a graph,
or network.  Upon calculating complex-network measures, we discover behaviors
traditionally associated with complexity.  We complement these complex-network
measures with complexity measures based on many-body quantities: persistent
entropy fluctuations and emergent phenomena.  This constellation of
observations, in the empirical and phenomenological spirit of complexity
science, leads to our conclusion that Goldilocks QCA produce physical
complexity.
}

Furthermore, we show that these Goldilocks rules can be implemented
with two classes of quantum computers: First, analog quantum
simulators~\cite{altman2019quantum} include Rydberg atoms in
optical lattices~\cite{weimer2010rydberg, bernien2017probing} and
trapped ions~\cite{blatt2012quantum, Zhang2017Nature}.  Second,
digital quantum computers include superconducting qubits, used
to demonstrate quantum supremacy~\cite{arute2019quantum}.  Quantum
simulators~\cite{buluta2009, Hauke2012, altman2019quantum} have energized the
developing field of nonequilibrium quantum dynamics~\cite{polkovnikov2011}. Our
work establishes a guiding question for this field: What are the origins of
complexity, % the features discoveredin multiscale complexity science, given
that we find complexity in simple, abiological quantum systems?

Our bridging of QCA with complexity science contrasts with all previous work on
QCA, to our knowledge
\footnote{
    The masters theses~\cite{hillberry2016entanglement, vargas2016quantum}
    defined QCA models presented here and considered the models' physical
    complexity. However, the present work reflects several hundred hours of
    additional compute time, new complexity measures, new analyses, and a
    refinement of the Goldilocks principle.
}.
For example, early QCA studies spotlighted single-particle or semiclassical
approximations~\cite{grossing1988quantum, lent1993quantum, meyer1996quantum,
Arrighi2019overview} or weak correlations at the level of density functional
theory~\cite{tougaw1996dynamic, farrelly2020review}. Other QCA work
concerns Clifford operators~\cite{Schlingemann_08_On, haah2018nontrivial,
freedman2019classification, freedman2019group, haah2019clifford}, which
can be efficiently simulated classically~\cite{Gottesman_98_Heisenberg}.
Also single iterations of QCA evolutions can be classically simulated
efficiently~\cite{Schumacher_04_Reversible, Gross_12_Index, Po_16_Chiral,
Gong_20_Classification,Sahinoglu_18_Matrix,Cirac_17_Matrix}.  In contrast,
our paper centers on emergent many-body phenomena. Additionally,
our methodology differs fundamentally from earlier works': Many of the
aforementioned papers are abstract and mathematical. We introduce into QCA
the toolkit of complexity science, which requires experimentation, empirical
observation, and data analysis.  While others have noted QCA-generated
entanglement~\cite{Bleh2012quantum, brennen2003entanglement, gutschow2010time,
Gutschow_10_Fractal, gopalakrishnan2018facilitated, gopalakrishnan2018operator,
vonkeyserlingk2018operator}, no previous work has bridged QCA to complexity
science.

In summary of our work's significance, we bridge the fields of quantum
computation and complexity science via QCA. QCA, we discover, can exhibit
physical complexity similar to that observed in biological, social, and
ecological systems. We quantify this complexity and prescribe how to define
QCA evolution rules that generate it.  Today’s Rydberg arrays, trapped ions,
and superconducting qubits can realize our QCA; hence our bridge between
disciplines is not only theoretical, but also experimentally accessible.

\edit{
This paper is organized as follows.  Section~\ref{sec:background}
introduces our QCA evolution schemes, including Goldilocks rules.
Section~\ref{sec:results} demonstrates our central claim: Goldilocks
QCA  generate physical complexity despite their simple unitary structure.
Sections~\ref{sec:background}-\ref{sec:results} suffice for grasping the
novelty of Goldilocks QCA's complexity generation.  Our QCA's  mathematical
details appear in Sec.~\ref{sec:models}.  In Sec.~\ref{sec:physical},
we propose a physical implementation of our QCA in Rydberg systems,
an  experimentally realized quantum simulator~\cite{altman2019quantum}.
Section~\ref{sec:conclusions} concludes with this work's significance.
}

\section{\label{sec:background}
    Overview of QCA}

\edit{
In this section, we first introduce classical elementary cellular automata
and compare them with our QCA (Sec.~\ref{sec:ECA_Intro}).  Second, we
establish the terminology and notation used to label and construct our QCA
(Sec.~\ref{sec:Define_QCA}).  For generality, and to enable simulation by
discrete and analog quantum computers, we define digital and continuous-time
QCA.  The number of QCA models is vast.  Therefore, we pare down the options
to a diverse and illustrative, but manageable, set (Sec.~\ref{sec:Scope}).
We identify the Goldilocks rules, which we contrast with trivial and
near-Goldilocks rules (Sec.~\ref{sec:Entree_Gold}).
}

\subsection{\label{sec:ECA_Intro}
    Classical elementary cellular automata}

\edit{
The best-known classical cellular automata are the 1D \emph{elementary cellular
automata} (ECA).  An ECA updates each bit of a length-$L$ string in discrete
time steps.  Each bit is updated in accordance with a local transition function
dependent on the bit's neighborhood.  The neighborhood is defined as the bit
and its two nearest neighbors.  The transition function is encoded in the
ECA \emph{rule number}, $C_R$. The $R = 0, 1, \ldots, 255$.  $C_R$ specifies
the bit's next state, given the neighborhood's current state.  Section
~\ref{sec:models} details the map from $R$ to the local transition function.
}

During each time step, all bits are updated simultaneously.  Such an update
requires a copy of the current bit string: Without the copy, one updates
the $j^{\rm th}$ bit, based on its neighbors, then must update the neighbors
based on the earlier $j^{\rm th}$ bit's value. But the earlier value has been
overwritten; hence the necessity of the copy.  Though simple to define, ECA
dynamics generate emergent features. Examples include fractal structures,
high-quality random number generation~\cite{wolfram1985cryptography},
and Turing-completeness (the ability to simulate any classical computer
program)~\cite{cook2004universality}.

Our QCA partially resemble, and partially contrast with, ECA.  Similarities
include one-dimensionality, spatial discreteness, and local updates' affecting
only a neighborhood's center site.  But we trade bits for qubits, securing new
features for, and restrictions on, our QCA.  First, time updates can entangle
qubits, generating quantum entropy absent from ECA.  Second, our QCA can run
in discrete or continuous time.  Hence digital and analog quantum computers
can implement our QCA.  Third, a QCA rule number specifies the neighborhood
conditions under which an operator evolves a qubit's state.  In contrast, an
ECA rule number specifies a bit's next state.  Fourth, our QCA evolve a qubit
under an operator independent of the qubit's state.  Our nearest-neighbor
QCA therefore parallel the 16 locally invertible (unitary) ECA.  Fifth,
in digital QCA, not all qubits are updated simultaneously.  As explained
above, a simultaneous update requires a copy of the current bit string, in
classical ECA.  So would a simultaneous update require a copy of the current
joint quantum state, in digital QCA.  The no-cloning theorem debars such a
copy
\edit{
    ~\cite{brennen2003entanglement, farrelly2020review}.
}
Finally, we generalize the local update rule to depend on $\geq 2$ neighbors,
as the ECA have been generalized~\cite{li1989complex}.

\subsection{\label{sec:Define_QCA}
    Defining quantum cellular automata}

\edit{
Section ~\ref{sec:models} will detail our QCA mathematically.  Here, we
provide the key terminology and notation.  $\hat{\sigma}^\alpha$ denotes the
$\alpha = x, y, z$ Pauli operator. The $\hat{\sigma}^z$ eigenbasis forms the
computational basis: $\ket{0}$ denotes the eigenstate associated with the
eigenvalue 1, and $\ket{1}$ denotes the eigenstate associated with $-1$.
}

\edit{
Many formal definitions of QCA are possible~\cite{farrelly2020review,
Arrighi2019overview}.  However, our definitions conform to the most general
notion of a QCA, as a dynamical quantum system that is causal, unitary, and
translation-invariant~\cite{Arrighi2019overview}
}
    \footnote{
        \edit{
        The translation-invariance requirement is often relaxed or
        reinterpreted, to accommodate QCA that are invariant under translations
        under just certain numbers of sites~\cite{Arrighi2019overview}.
        Also, nonunitary QCA form an emerging branch of
        study~\cite{brennen2003entanglement, farrelly2020review,
        wintermantel2020unitary}.
        }
    }.
\edit{
Furthermore, our family of QCA overlaps with the family defined
in~\cite{brennen2003entanglement, wintermantel2020unitary}. The
overall structure of their QCA definition resembles the structure
of our definition. Differences include our QCA’s accommodating
five-site neighborhoods and their QCA’s accommodating nonunitary
evolutions. Additionally, our family of analog QCA includes, as a special case,
the QCA in~\cite{Bleh2012quantum}.
}

A QCA updates each qubit of a length-$L$ chain according to a unitary update
rule. Qubit $j$ has a set $\Omega_j$ of \emph{neighbors}: the sites, excluding
$j$, within a radius $r$ of $j$.  $\Omega_j$ corresponds to a Hilbert space
on which are defined density operators that form a vector space spanned
by projectors $\op{\mathcal{P}}^{(i)}_{\Omega_j}$. The $i$ enumerates the
neighbors' possible configurations, and the projectors form a complete set:
$\sum_i \op{\mathcal{P}}^{(i)}_{\Omega_j} = \op{\Id}$.  The \emph{neighborhood}
of site $j$ consists of $j$ and its neighbors.

We define digital time evolution as follows.  Given an ECA rule number, we
convert into a unitary that acts on a qubit's neighborhood.  The gate evolves
different neighborhoods in different circuit layers.
\edit{
    This layered evolution resembles the Trotterization of a many-body Hamiltonian for simulation on a digital quantum computer.
}

The unitary on site $j$'s neighborhood has the form
\begin{equation} \label{eq:QCAdig}
    \op{U}_j := \sum_{i} \op{V}^{c_i}_j \op{\mathcal{P}}^{(i)}_{\Omega_j}\, .
\end{equation}
$\op{V}_j := \Id^{\otimes j} \otimes \op{V} \otimes \Id^{\otimes (L -
1 - j)}$ denotes a unitary activation operator $\op{V}$ that nontrivially
evolves site $j$.  The rule number's binary expansion determines the $c_i \in
\{0,1\}$.  The map between rule and $\{ c_i \}$ is tedious, so we relegate it
to Sec.~\ref{sec:Def_T}.  The matrix power implies that $\op{V}^{c_i} \in \{
\op{\Id}, \op{V} \}$. \edit{If $\hat{V} = \hat{\sigma}^{x}$ and the initial
state is a product of $\ket{0}$'s and $\ket{1}$'s, the dynamics are classical.
We therefore omit this $\hat{V}$ from this paper.}

To define analog time evolution, we convert the rule number into a many-body
Hamiltonian $\sum_j \op{H}_j$. The single-neighborhood Hamiltonian is defined
as
\begin{equation} \label{eq:QCAana}
    \op{H}_j := \op{h}_j  \sum_i c_{i} \op{\mathcal{P}}^{(i)}_{\Omega_j} \, .
\end{equation}
$\op{h}_j$ denotes a Hermitian activation operator that acts on site $j$.
The rule number's binary expansion determines the $c_i \in \{0,1\}$.

\edit{
To evolve the qubits under an analog QCA rule, we exponentiate the many-body
Hamiltonian, forming $\op{U} = \exp(-i \delta t \op{H})$.  This $\op{U}$
evolves the system for a time $\delta t$. Thus, evolving for one time unit
requires $1/\delta t$ applications of $\op{U}$.  Choosing $\delta t \ll 1$,
we approximate $\op{U}$ with a product of local unitaries to order $\delta
t^4$~\cite{nielsen2010quantum}
}.

Analog and digital QCA tend to yield qualitatively similar outcomes. The
relationship between analog and digital QCA is formalized in
App.~\ref{sec:analog_digital_relationship}.

\subsection{\label{sec:Scope}
    Scope}

\edit{
Our digital QCA~\eqref{eq:QCAdig} and analog QCA~\eqref{eq:QCAana} offer
rich seams to explore: Many neighborhood sizes, time-evolution schemes,
activation operators, boundary conditions, etc. are possible.  One paper
cannot report on all the combinations, so we balance breadth with depth.
Below, we identify a manageable but representative subset of options.
Appendix~\ref{sec:supplemental} demonstrates that conclusions of ours
generalize to other initial conditions, boundary conditions, and activation
operators.
}

\edit{
We illustrate digital QCA with $r=1$, or nearest-neighbor, rules.  This choice
facilitates comparisons with ECA.  We denote $r=1$ digital QCA by $T_R$.
The $T$ evokes three-site neighborhoods; and $R = 0, 1, \ldots, 15$ denotes
the rule number.  We illustrate $T_R$ rules with a Hadamard activation
operator, $\op{V} = \subs{\op{U}}{H} := ( \op{\sigma}^{z} + \op{\sigma}^{x}
) / \sqrt{2}$, which interchanges the Bloch sphere's $x$- and $z$-axes.
We choose the Hadamard so that the QCA generate entanglement even when the
lattice begins in a computational-basis state.  During each time step, even-$j$
sites are updated via the unitary~\eqref{eq:QCAdig}, followed by odd-$j$ sites
[Fig.~\ref{fig1}(e)].  We showcase these digital rules due to their conceptual
simplicity, ease of implementation, and similarities with ECA.
}

\edit{
We illustrate analog QCA with a time resolution of $\delta t = 0.1$ and $r =
2$, i.e., 5-site neighborhoods,
    \footnote{
        Five qubits form the smallest neighborhood for which a
        Goldilocks rule generates entangled breathers. Such breathers
        are emergent phenomena indicative of complexity and discussed in
        Sec.~\ref{sec:results_breather}.
    }.
As $6.55\times 10^{4}$ such rules exist, we must pare them down.  We do so by
focusing on \emph{totalistic} rules, which update a site conditionally on the
total number of neighbors in $\ket{1}$.  Only 32 totalistic $r=2$ rules  exist.
Furthermore, totalism endows left-hand and right-hand neighbors with the
same influence, enforcing a physically common inversion symmetry.  We denote
these rules by $F_R$.  The $F$ refers to five-site neighborhoods; and $R = 0,
1, \ldots, 31$ denotes the totalistic-rule number.  We choose the activation
$\op{h}_j=\op{\sigma}_j^x$, for three reasons. First, $\op{\sigma}_j^x$ serves
as the Hamiltonian in many quantum simulations. Second, $\op{\sigma}_j^x$
transforms the computational basis, and so our typical initial states,
nontrivially. Third, $\op{\sigma}_j^x$ can generate entanglement when serving
as a Hamiltonian $\op{h}_j$ (as opposed to serving as a unitary $\op{V}_j$).
}

\edit{
In all simulations reported on in the main text, the boundary conditions
are fixed: Consider evolving a boundary site, or any of the boundary's $r$
neighbors. The evolution is dictated partially by imaginary sites, lying
past the boundary, fixed to $\ket{0}$'s.  Again, we exhibit other boundary
conditions in App.~\ref{sec:supplemental}, supporting the generality of the
main text's conclusions.
}

\edit{
Throughout the main text, $T_R$ simulations are initialized with
the product state $\ket{\ldots 000 \: 1 \: 000\ldots}$.  This choice
promotes simplicity; concreteness; and ease of comparison with ECA, which
are often initialized similarly.  $F_R$ simulations are initialized in
$\ket{\ldots 000 \: 101 \: 000 \ldots}$: One neighborhood must contain at
least two $\ket{1}$'s for the Goldilocks $F_R$ rule to evolve the state
nontrivially, as we will see below.  Furthermore, the Goldilocks $F_R$
rule evolves this state into an \emph{entangled breather}.  This breather,
first shown in Sec.~\ref{sec:results_one-point} and further analyzed in
Sec.~\ref{sec:results_breather}, parallels the breather of discrete nonlinear
wave theory~\cite{flach2008discrete, lederer2008discrete}.
}

\subsection{\label{sec:Entree_Gold}
    Entr\'{e}e into Goldilocks and non-Goldilocks rules}

\edit{
To gain intuition about Goldilocks rules, consider the extreme digital rules
$T_0$ and $T_{15}$.  $T_0$ always applies the identity operator to the target
site.  $T_{15}$ applies the activation operator to all sites at every time
step.  Both dynamics are trivial and generate no entanglement.  As Goldilocks
would say, $T_0$ is too passive, and $T_{15}$ is too active.
}

\edit{
Interpolating between the extremes yields rules that increasingly approach what
Goldilocks would deem ``just right.''  Consider extending the do-nothing rule
$T_0$ to updating sites whose two neighbors are in $\ket{0}$.  The resulting
rule, $T_1$, leads to nontrival dynamics.  Similarly, consider constraining the
update-everything rule $T_{15}$: Each site is updated unless both its neighbors
are in $\ket{0}$'s.  Nontrivial dynamics emerge from the resulting rule,
$T_{14}$.
}

\edit{
Optimizing the tradeoff between passivity and activity leads to the Goldilocks
rule $T_6$.  Similar tradeoffs have been observed to generate and maintain
complexity~\cite{carlson2002complexity}.  Section~\ref{sec:results} shows that
tradeoffs generate complexity also in quantum dynamics and computation.
}

Let us examine the form of $T_6$.  Let $\op{P}^{(m)}=\proj{m}$ denote a
projector onto the $\ket{m}$ eigenspace of $\hat{\sigma}^z$, for $m = 0, 1$.
The $T_6$ local update unitary, acting with a general activation operator, is
\begin{align}
   \label{eq_T_6_U}
   \op{U} & =\op{P}^{(1)}\otimes\op{\Id}\otimes\op{P}^{(1)}
   + \op{P}^{(1)}\otimes\op{V}\otimes\op{P}^{(0)}
   \nonumber \\ & \quad
   + \op{P}^{(0)}\otimes\op{V}\otimes\op{P}^{(1)}
   + \op{P}^{(0)}\otimes\op{\Id}\otimes\op{P}^{(0)} .
\end{align}
The Hadamard acts if exactly one neighbor is in $\ket{1}$---not 0 (too few)
neighbors or 2 (too many) neighbors.

\edit{
Having introduced the digital Goldilocks rule, we define the analog Goldilocks
rule, $F_4$.  $F_4$ updates a site if exactly 2 of the site's 4 neighbors
are in $\ket{1}$.  Let us construct the single-neighborhood Hamiltonian,
$\op{H}_j$, for a Hermitian  activation $\op{h}_j$.  We sum over the ways of
distributing two $\op{P}^{(1)}$'s across four neighbors:
\begin{align}
   \op{H}_j = \op{h}_j
   \sum_{ \substack{ \alpha,\beta,\gamma,\delta \in \{0,1\}, \\
                     \alpha+\beta+\gamma+\delta = 2 }}
   \op{P}^{(\alpha)}_{j-2} \;
   \op{P}^{(\beta)}_{j-1} \;
   \op{P}^{(\gamma)}_{j+1} \;
   \op{P}^{(\delta)}_{j+2} \; .
\end{align}
}

\edit{
Having introduced the Goldilocks rules, we identify a near-Goldilocks rule that
has been realized experimentally but whose status as a QCA has not been widely
recognized.  The PXP model is a Hamiltonian that has been implemented with
Rydberg-excited neutral atoms~\cite{bernien2017probing} [Fig.~\ref{fig1}(d)].
Such models generate diverse quantum dynamical features, such as many-body
scars~\cite{turner2018quantum}.  Studies of PXP have spotlighted approaches
conventional in many-body physics, such as energy-level statistics. Any
physical complexity generated by the model, and the model's standing within
the broader class of QCA, have not been explored. We quantify the complexity
and identify the model as a QCA, although PXP is not a focus of this paper: As
shown below, PXP (more precisely, a digital version of PXP) is a particular
case of the near-Goldilocks rule $T_1$. The model therefore generates
less complexity than Goldilocks rules, our main focus.  Additionally,
rule $T_{14}$ has been explored theoretically under the name ``rule
54''~\cite{alba2019operator, lesanovsky2019non}.  We contextualize $T_{14}$
and $T_1$, demonstrating that they are near-Goldilocks rules that belong to a
broader family of QCA, which contains QCA that produce greater complexity.
}

\edit{
To demonstrate the the power of this cellular-automaton-and-complexity-driven
approach, we arrive at the PXP model not via conventional many-body
considerations, but via ECA; and we observe rich many-body physics accessible
to QCA beyond PXP: Figure~\ref{fig1}(a) features the ECA rule $C_{201}$, known
to produce a simple periodic pattern called a \emph{blinker}.  The name derives
from the oscillations of black and white squares that represent 1 and 0 bits.
The closest digital QCA rule is $T_1$, of which the digital PXP model is a
particular case, which generates Figs.~\ref{fig1}(b)-(c).  A quantum analog
of a bit's being 1 or 0 is $\langle \hat{\sigma}^z \rangle$.  $T_1$ produces
not only visually richer dynamics than the ECA, but also quantum entropy.
Progressing beyond PXP to other QCA reveals a truly nonclassical analog of a
blinker, we show in the next section.
}

\section{\label{sec:results}
    Main results: Goldilocks QCA generate complexity}

\edit{
The prevalence of tradeoffs in complexity, with Goldilocks rules'
regulation of a tradeoff, suggests that Goldilocks rules may generate
complexity.  This section supports this expectation with numerical
evidence.  As many readers may be unfamiliar with complexity science,
Sec.~\ref{sec:results_one-point} initiates our numerical study of
Goldilocks rules with measures familiar from many-body physics: one-point
correlation functions and von Neumann entropies.  These metrics equilibrate
under non-Goldilocks evolution but not under Goldilocks evolution.
Building on this observation of Goldilocks rules' noteworthy dynamics,
we calculate four complex-network measures~\cite{Valdez2017Quantifying,
bagrov2020detecting, walschaers2020emergent, sundar2020response} in
Sec.~\ref{sec:results_network-complexity}.  We evaluate these measures on
networks formed from the mutual information between qubit pairs, in states
generated by QCA.  We complement these two-point functions with many-body
entropy fluctuations in Sec.~\ref{sec:results_bond-entropy}.  Additionally,
the Goldilocks rule $F_4$ exhibits an emergent feature, the nonclassical analog
of the ECA blinker. We study the structure and robustness of the entangled
breather in Sec.~\ref{sec:results_breather}.  As explained in the introduction,
complexity science is phenomenological.  A system is regarded as complex
if it exhibits several features suggestive of complexity~\cite{Newman2003}.
Together, the network measures, entropy fluctuations, and entangled breather
form six signatures of complexity.  We therefore conclude that Goldilocks QCA
generate complexity.
}

\subsection{\label{sec:results_one-point}
    One-point measures}

\begin{figure*}[t]
\begin{centering}
\includegraphics[width=0.85\textwidth]{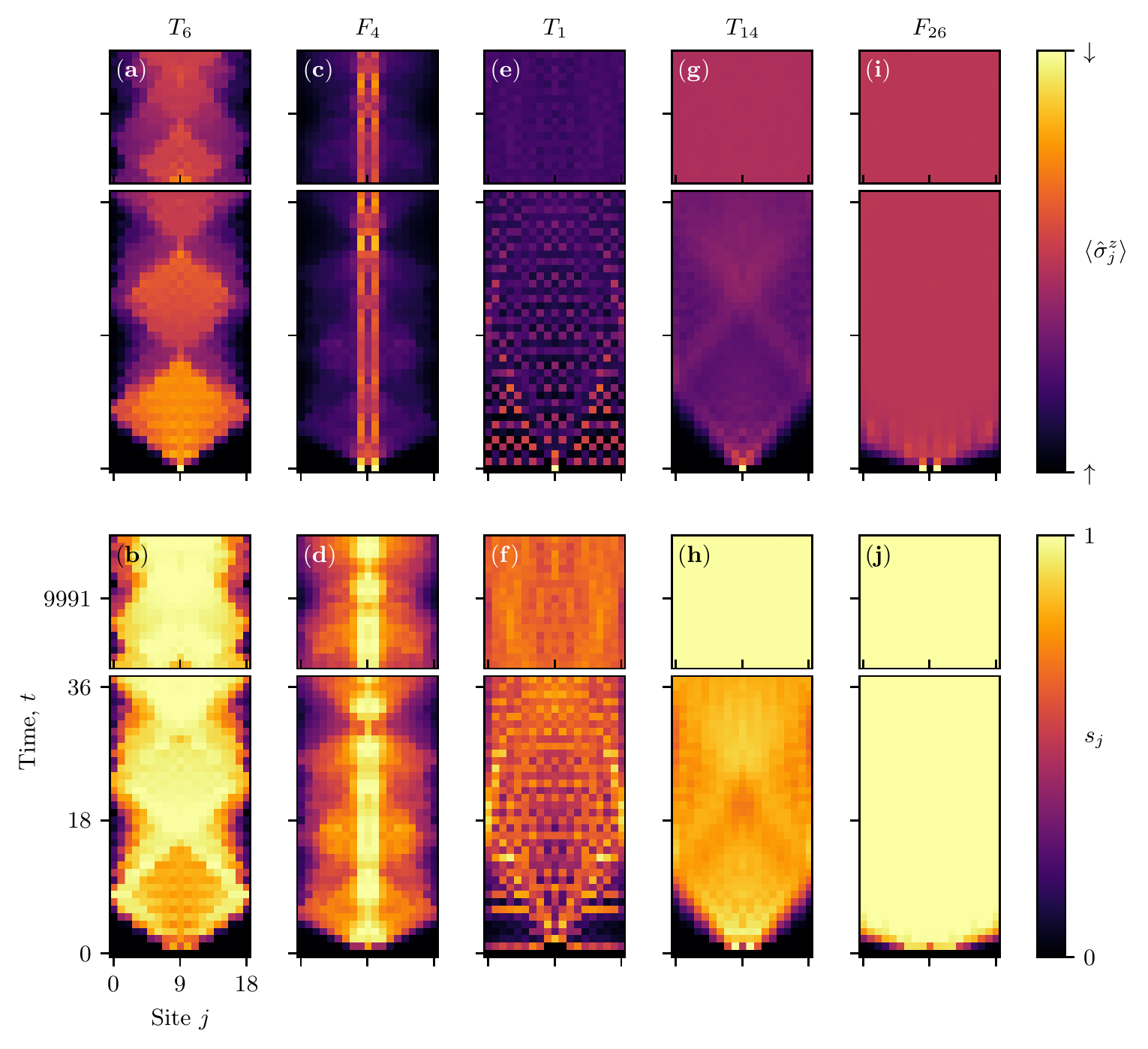}
\caption{
    QCA produce diverse dynamical behaviors, visualized here with familiar
    one-point observables.  For (a)-(b) and (e)-(h), the digital $T_R$
    rules, an $L{=}19$-site chain was initialized to $\ket{\ldots 000 \: 1
    \: 000\ldots}$, then evolved in discrete time under various $T_R$ rules.
    The Goldilocks rule $T_6$ generates the spin dynamics shown in (a) and the
    entropy  dynamics shown in (b). The dynamics exhibit patterns that resist
    equilibration for long times, as shown in each panel's upper block.
    For (c)-(d) and (i)-(j), we extend QCA to 5-site rules, initialize the
    $L{=}19$-site chain to $\ket{\ldots 000 \: 101 \: 000\ldots}$, and evolve
    in continuous time.
    (c)-(d) The Goldilocks rule $F_4$ generates a nonclassical variation on the
            blinker in Fig.~\ref{fig1}(a), an \emph{entangled breather}.
    (i)-(j) In contrast, far-from-Goldilocks rule $F_{26}$ promotes rapid
            equilibration to a high-entropy state.
    \label{fig2}
    }
\end{centering}
\end{figure*}

\edit{
As complexity science may be unfamiliar to many readers, our numerical study
of Goldilocks rules begins with measures familiar from many-body physics.
Figure~\ref{fig2} shows $\braket{\op{\sigma}_j^z}$ (top row) and the
single-site von Neumann entropy $s_j$ (bottom row) as functions of time.  The
entropy of a subsystem $\mathcal{A}$ in a reduced state $\rho_{\mathcal{A}}$
is defined as $s_{\mathcal{A}} := - {\rm Tr} (\rho_{\mathcal{A}} \log
\rho_{\mathcal{A}})$.  Further definitions and calculations appear in
Apps.~\ref{sec:expval} and~\ref{sec:entropy}.
}

\edit{
Under the QCA evolutions illustrated in the right-hand panels,
$\braket{\op{\sigma}_j^z}$ and $s_j$ equilibrate: After a time of the
order of or much longer than the propagation time across the system, the
plots become fairly uniform across space and show only small fluctuations.
The Goldilocks rules, $T_6$ and $F_4$, defy equilibration most dramatically
[Figs.~\ref{fig2}(a)-(b) and (c)-(d)].  Their measures display persist
patterns.  For instance, $F_4$ generates an entangled breather analyzed in
Sec.~\ref{sec:results_breather}.
}

\edit{
The near-Goldilocks rule $T_1$ [Figs.~\ref{fig2}(e)-(f)] resists equilibration
second-best: Although its patterns fade, they persist across time.  The fading
is stronger under the near-Goldilocks rule $T_{14}$ [Figs.~\ref{fig2}(g)-(h)].
Rule $F_{26}$ [Fig.~\ref{fig2}(i) and (j)] encourages equilibration the most:
$\braket{\op{\sigma}_j^z}$ quickly equilibrates to zero (rose color), and $s_j$
quickly equilibrates to yellow (a high entropy value) across the lattice.
$F_{26}$ is a far-from-Goldilocks rule: It evolves site $j$ if the site has
one, three, or four neighbors in $\ket{1}$.
}

\edit{
We illustrate far-from-Goldilocks rules with $F_{26}$, Goldilocks rules with
$T_6$ and $F_4$, and near-Goldilocks rules with $T_1$ and $T_{14}$ throughout
the main text.  Appendix~\ref{sec:supplemental} widens the survey to rule
$T_{13}$, a far-from-Goldilocks, nontotalistic digital  rule.
}

\edit{
One might wonder whether integrability or finite-size effects prevent
the Goldilocks rules from equilibrating the measures in Fig.~\ref{fig2}.
Finite-size effects cannot take responsibility because the non-Goldilocks rules
promote equilibration at the same system sizes used for the Goldilocks rules.
Furthermore, App.~\ref{sec:supplemental} demonstrates our results' robustness
with respect to changes in boundary conditions.  Whether Goldilocks rules are
integrable is discussed in Sec.~\ref{sec:conclusions} but is irrelevant to
the present paper, which focuses on complexity.  The important point is that
Goldilocks rules differ from other QCA even according to many-body measures. We
now show that complexity distinguishes Goldilocks rules.
}

\subsection{\label{sec:results_network-complexity}
    Complexity measures evaluated on mutual-information networks}

\begin{figure*}[t]
\begin{centering}
\includegraphics{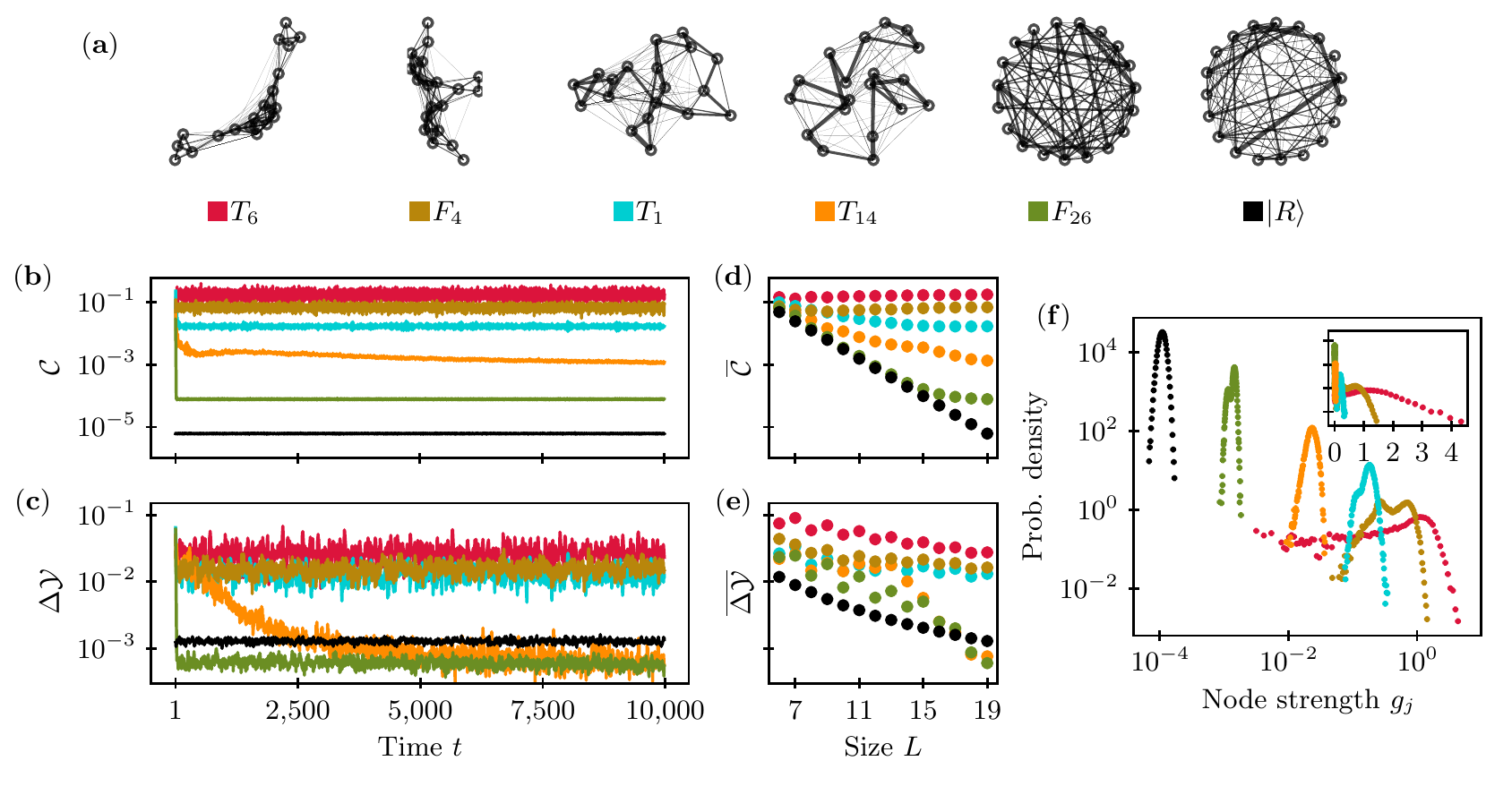}
\caption{\label{fig3}
    \edit{
    Mutual information networks distinguish Goldilocks rules as generating the most complexity.
    (a) An $L{=}19$-site spin chain was initialized, then evolved for 500
        time units.  In each of three cases, the initialization was to
        $\ket{0 0 \ldots 0 \, 1 \, 0 0 \ldots 0}$, and a $T_R$ QCA dictated
        the evolution.  In two other cases, the initialization was to $\ket{0
        0 \ldots 0 \, 1 0 1 \, 0 0 \ldots 0}$, and an $F_R$ QCA dictated the
        evolution.  The open circles represent the qubits.  Nodes $j$ and $k$
        are connected by a line whose thickness is proportional to the mutual
        information $M_{jk}$. The nodes' layout is found by modeling links
        as springs; well-connected nodes lie close together.  % The initial
        conditions are chosen to match those of Fig.~\ref{fig2}.  The networks'
        labels provide a legend for the panels below.  The black lines and
        dots show the results of initializing the qubits to a random state
        $\ket{R}$, evolving the state under one of the five QCA, then averaging
        the final state over the QCA.
    (b) Clustering coefficient $\mathcal{C}$ vs. time $t$.
    (c) Disparity fluctuations $\Delta \mathcal{Y}$ vs. $t$.
    (d)-(e) Clustering coefficient $\mathcal{C}$ and disparity fluctuations
            $\Delta \mathcal{Y}$, averaged over late times $t \in [5,000,~10,000]$
            (denoted by the overline), vs. system size $L$.
    (f) Probability density over node strengths $g_j$ collected at $t \in [5,000,~
        10,000]$. Statistics were accumulated with logarithmically spaced bins. Only bins
        with $>75$ counts were maintained. The inset shows the same data on a
        linear horizontal scale and identical vertical scale.}
    }
\end{centering}
\end{figure*}

\edit{
We quantify QCA's complexity using complex networks.  A \emph{network}, or
graph, is a collection of nodes, or vertices, connected by links, or edges.
A \emph{complex network} is neither regular nor random.  Completely connected
graphs (in which all nodes link to all others) are not complex; their links
are regular and orderly. For the same reason, lattice graphs are not complex
networks. Erd\"{o}s-R\'{e}nyi graphs is not complex for the opposite reason:
Their links are created randomly.
}

\edit{
Complex networks lie between the random and regular extremes.  Concrete
examples include social, airport-connection, neurological, and metabolic
networks. Abstract examples include  small-world networks, which exhibit
clustering (nodes collect into tight-knit groups) and a short average path
length (one node can be reached from another via few links, on average).
}

\edit{
We study networks that emerge from quantum many-body systems.
Networks can alternatively be imposed on quantum systems, such as
networked NISQ (noisy intermediate-scale quantum) computers and a quantum
Internet~\cite{awschalom2019development, Biamonte_19_Complex}.  Yet complex
networks can also emerge naturally in quantum states generated by noncomplex
systems or models.  This emergence has been demonstrated with the Ising and
Bose-Hubbard models near quantum critical points~\cite{Valdez2017Quantifying,
Sundar2018Complex}.  We progress beyond critical phenomena to dynamical quantum
systems and quantum computation.
}

\edit{
Networks have been formed from the mutual information, as in brain activity
that exhibits a small-world structure~\cite{jin2012reorganization,
Bullmore2009}.  The mutual information quantifies the correlation
between two members of a system.  We use the quantum mutual
information~\cite{detomasi2017quantum, herviou2019multiscale} between qubits
$j$ and $k \neq j$,
\begin{align}
   M_{jk} := \frac{1}{2} ( s_j + s_k - s_{jk} ) \, .
\end{align}
The $1/2$, though nonstandard, conveniently normalizes the quantity as
$M_{jk} \leq 1$.  The mutual information depends on two-site von Neumann
entropies that can be measured experimentally with quantum overlapping
tomography~\cite{Cotler_2020_quantum}.  $M_{jk}$ upper-bounds every
same-time, two-point correlator between $j$ and $k$~\cite{wolf2008area}.
Appendix~\ref{sec:entropy} contains further details.  We denote by $M$ the
matrix of elements $M_{jk}$.
}

\edit{
Figure~\ref{fig3}(a) displays mutual-information networks formed by QCA after
500 time units.  Open circles represent the qubits, which serve as nodes.
The greater the correlation $M_{jk}$, the closer together nodes $j$ and $k$
lie, and the thicker the line between them.  The Goldilocks rules $T_6$ and
$F_4$ produce visibly clustered networks.  The non-Goldilocks rules' networks
more closely resemble the network encoded in the random state $\ket{R}$.
$\ket{R}$ is an $L$-qubit state of $2^L$ complex amplitudes.  The amplitudes'
real parts are drawn from independent, identical Gaussian distributions,
as are the imaginary parts.  The near-Goldilocks rules $T_1$ and $T_{14}$
generate networks that interpolate between the Goldilocks and random networks.
The far-from-Goldilocks rule $F_{26}$ produces a network strikingly similar to
$\ket{R}$'s in shape and connectivity. Progressing beyond these qualitative
features, we now quantify the networks' complexity with four metrics:
clustering, average path length, disparity fluctuations, and the probability
density over node strengths.
}

\subsubsection{Clustering and average path length}

\edit{
The \emph{clustering coefficient} $\mathcal{C}$ quantifies transitivity,
easily intuited in the context of social networks: If Alice is friends with
Bob and friends with Charlie, how likely is Bob to be friends with Charlie?
The clustering coefficient is proportional to the number of connected triangles
in the network, divided by the number of possible connected triangles:
\begin{equation} \label{eq:clustering}
    \mathcal{C} := \frac{\Tr \left (M^3 \right ) }
    {\sum_{\substack{j,k=0 \\ j \neq k}}^{L-1}\left [M^2 \right ]_{jk}}
    \in [0, 1] .
\end{equation}
We will measure the clustering as a function of time [Fig.~\ref{fig3}(b)] and
system size [Fig.~\ref{fig3}(d)].
}

\edit{
Figure~\ref{fig3}(b) shows the clustering coefficient averaged over the
network, $\mathcal{C}$, as a function of time.  Complex networks exhibit large
clusterings, as discussed above. The Goldilocks rules (ruby and amber curves)
maintain the highest clusterings.  The nearest clustering characterizes the
near-Goldilocks rule $T_1$ (turquoise curve). The $T_6$ clustering (ruby curve)
exceeds the $T_1$ clustering by a factor of 10.5, and the $F_4$ clustering
(amber curve) does by a factor of 2.2. The lowest clustering appears in the
random network (black curve), generated as follows: The lattice is initialized
in a random state $\ket{R}$, then evolved under one of the five QCA. The
clustering is measured, then averaged over the five QCA.
}

\edit{
The qualitative trends remain robust as the system size, $L$, grows:
Figure~\ref{fig3}(d) shows the clustering, averaged over the network and over
late times $t \in [5,000,~10,000]$ as a function of $L$.  Only under the Goldilocks
rules $T_6$ and $F_4$ (ruby and amber dots) does $\mathcal{C}$ grow with $L$.
Under the other QCA, $\mathcal{C}$ decays exponentially, then plateaus.  These
behaviors approximate the behavior in the random state, wherein $\mathcal{C}$
strictly decays.  Hence only the Goldilocks rules' complexity, as quantified
with clustering, remains robust with respect to system-size growth.
}

\edit{
As mentioned above, clustering combines with short average paths in
small-worlds networks, which are complex.  All our QCA simulations generated
networks with short average path lengths $\sim 1.5$
    \footnote{
        \edit{
            Path length is defined straightforwardly for unweighted networks,
            whose links all represent 1s in an adjacency matrix.  Our networks
            are weighted; the links represent real numbers $M_{jk} \in [0,
            1]$. In accordance with convention~\cite{jin2012reorganization},
            we introduced a threshold $q$: All $M_{jk} < q$ are mapped
            to 0, while all $M_{jk} \geq q$ are mapped to 1. We checked
            thresholds $q$ below, at, and above the median link weight.
            Conventionally~\cite{reka2002statistical, Newman2003}, a path
            length is considered short if it grows logarithmically in the
            network size.  When $L = 19$, $\ln L = 2.94$, which, for our
            choices of $q$, exceeds the average path length of $\sim1.5$ that
            we measured.
        }
    }.
Since Goldilocks rules generate high clustering and short average paths,
Goldilocks rules generate small-world networks and so complexity.
}

\subsubsection{Disparity fluctuations}

\edit{
Having measured the clustering and average path length, we turn to disparity.
The \emph{disparity} quantifies how heterogeneous a network is, or how much the
network resembles a backbone
    \footnote{
        \edit{
            In some cases, the network is nearly disconnected:
            $M_{jk} \approx 0$.  Such disconnection poses numerical
            challenges.  We promote the calculations' convergence by adding
            a small imaginary part $\sim 10^{-16}$ to the denominator
            of Eq.~\ref{eq:disparity}, then taking the real part of the
            result~\cite{vargas2016quantum}.
        }
    }:
\begin{equation} \label{eq:disparity}
    \mathcal{Y} := \frac{1}{L} \sum_{j=0}^{L-1}
        \frac{\sum_{k=0}^{L-1}\left ( {M_{jk}} \right )^2}
        {\left ( \sum_{k=0}^{L-1} M_{jk}\right )^2}
    \in [0, 1) \, .
\end{equation}
Technically, $\mathcal{Y}$ is the disparity averaged over the network; the
disparity is conventionally defined on one site. However, we call $\mathcal{Y}$
``the disparity'' for conciseness.
}

\edit{
Two examples illuminate this definition.  First, consider a uniform network:
$M_{jk} = a$ is constant for all $j \neq k$. Along the diagonal, $M_{jj}=0$.
The uniform network lacks a backbone and so has a small disparity: If $L > 1$
denotes the number of nodes, $\mathcal{Y}=1 / (L-1) \to 0$ as $L \to \infty$.
In contrast, a 1D chain has an adjacency matrix $M_{jk} = a (\delta_{j (k+1)}
+ \delta_{j (k-1)} )$.  The chain consists entirely of a backbone, so the
disparity is order-one for all system sizes $L$: $\mathcal{Y}=(L+2)/(2L)$.
Disparity has been applied in complex-network theory, e.g., to classify
chemical reactions in bacterial metabolism~\cite{almaas2004global}.
}

\edit{
Figure~\ref{fig3}(c) shows fluctuations $\Delta \mathcal{Y}$ in QCA disparity
as a function of time.  We define $\Delta \mathcal{Y}$ as the standard
deviation of $\mathcal{Y}$ in a rolling time window of $L$ time units.  $\Delta
\mathcal{Y}$ reflects changes in a network's uniformity.  The Goldilocks rule
$T_6$ (ruby line) always produces disparity fluctuations above the pack's.
The Goldilocks rule $F_4$ (amber line) always remains at the top of the
pack, though the near-Goldilocks rules ($T_1$ and $T_{14}$, represented with
turquoise and yellow lines) do, too.  The far-from-Goldilocks rule $F_{26}$
(green line), like the random state $\ket{R}$ (black line), produce 1-2 orders
of magnitude fewer disparity fluctuations.  The $\ket{R}$ data are generated
analogously to in the clustering analysis.
}

\edit{
Similar trends surface in Fig.~\ref{fig3}(e), which shows disparity
fluctuations averaged over space and over late times, $\overline{\Delta
\mathcal{Y}}$, as a function of system size.  The $\overline{\Delta
\mathcal{Y}}$ produced by the Goldilocks rules (ruby and amber dots) exceed the
$\overline{\Delta \mathcal{Y}}$ produced by the other QCA, at all system sizes.
Again, the near-Goldilocks rules (turquoise and yellow) mimic the Goldilocks
rules most, and the far-from-Goldilocks rule (green) mimics the random state
most (black).  High disparity fluctuations imply a lack of equilibration in
network structure: The network sometimes exhibits heterogeneity and sometimes
exhibits uniformity.  Such diversity signals the complexity of, e.g., Internet
traffic~\cite{duch2006scaling}. Here, the diversity signals the complexity of
Goldilocks and, to some extent, near-Goldilocks rules.
}

\subsubsection{Probability density over node strengths}

\edit{
Our fourth complex-network measure is the probability density over node
strengths.  The \emph{node strength}, $g_j:=\Sigma_k M_{jk} \geq 0$, measures
how strongly node $j$ is connected to the rest of the network.  We draw this
measure from analyses of social and airport-connection networks, real-world
complex networks~\cite{barrat2004architecture, antoniou2008statistical}. These
networks contain \emph{hubs}: Few nodes carry a significant fraction of the
network's total node strength. Hubs make the the node-strength distribution
broad, e.g., power-law tails.
}

\edit{
Figure~\ref{fig3}(f) displays the node-strength densities of the networks
generated by the QCA.  The probability densities were calculated at late times
$t \in [5,000,~10,000]$.  The Goldilocks rules ($T_6$ and $F_4$, shown in ruby and
amber) produce distributions that are significantly broader than the others:
The $T_6$ points form a lopsided peak with a fat tail.  The $F_4$ points form
two peaks that ensure the distribution's width.  Hence Goldilocks rules cause a
few nodes to be strongly connected, as in known complex networks.  The other
rules, and the random state, lead to narrowly peaked distributions: Most
connection strengths are centered on the mean.
}

\edit{
We have investigated four complexity measures on mutual-information networks
generated by QCA: clustering, average path length, disparity fluctuations,
and the probability density over node strengths. The Goldilocks rules tend to
exhibit the most complexity; the near-Goldilocks rules, the second-most; the
far-from-Goldilocks rule, less; and the random state, the least.
}

\subsection{\label{sec:results_bond-entropy}
    Bond entropy and its fluctuations}

\begin{figure*}[ht]
\begin{centering}
\includegraphics[width=1.0\textwidth]{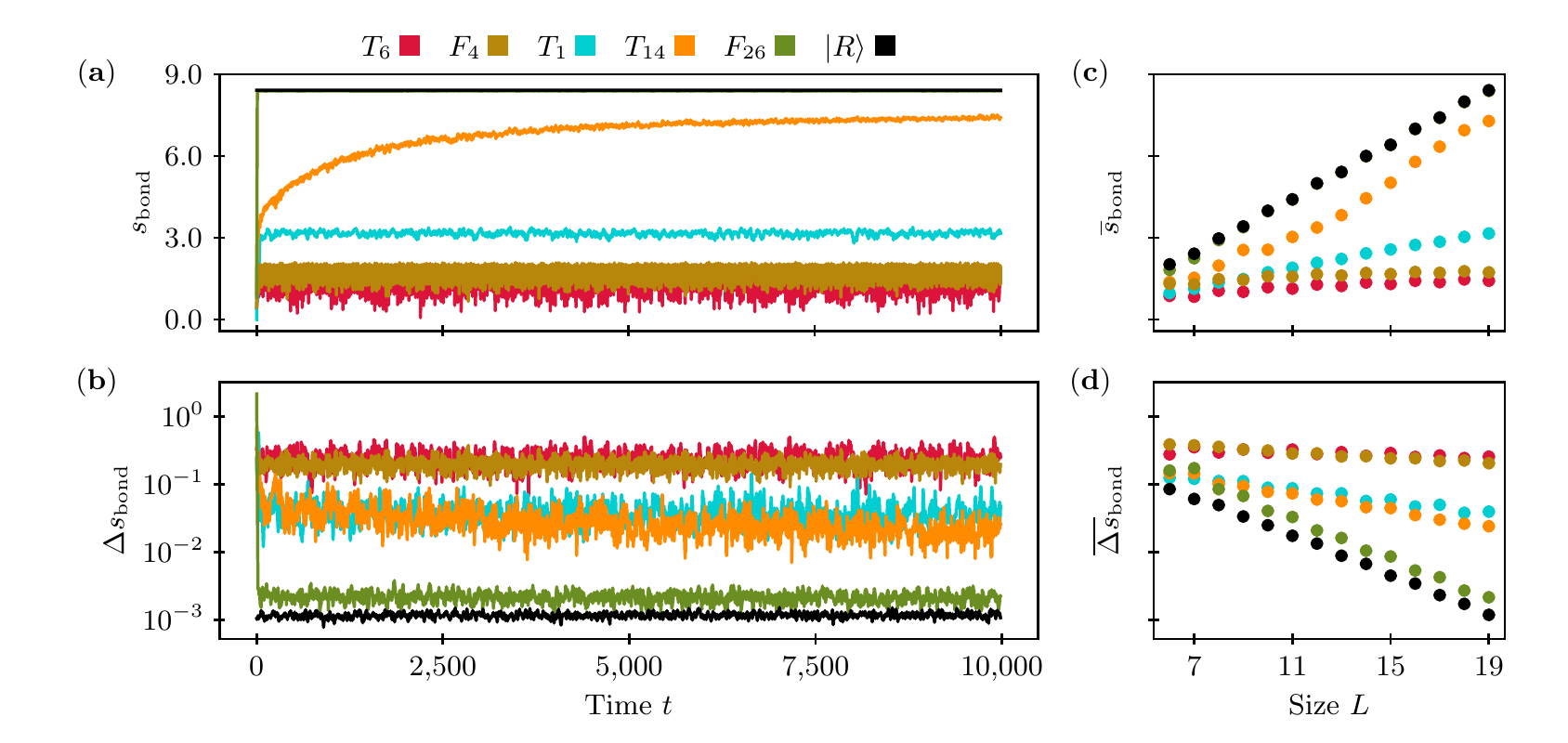}
\caption{\label{fig4}
    \edit{
    Entropy, and fluctuations in the entropy, of the qubit chain's central
    bond. Large entropy fluctuations are characteristic of macro-scale
    complex systems. Goldilocks rules display the largest such entropy
    entropy fluctuations of all the rules considered here.  The black lines
    and dots, labeled $\ket{R}$, were generated as described in the caption
    of Fig.~\ref{fig3}.  The digital $T_R$ QCA and analog $F_R$ QCA were
    initialized as described there.
    (a) The central cut's second-order R\'enyi bond entropy, $\subs{s}{bond}$,
        as a function of time for systems of size L{=}19.
    (b) The bond entropy averaged over late times $t \in [5,000,~10,000]$,
        $\subs{\overline{s}}{bond}$, as a function of system size.
    (c) Entropy fluctuations $\Delta \subs{s}{bond}$, computed as a rolling
        standard deviation with a window size of $L{=}19$ time units.
    (d) Time-averaged fluctuations in the entropy, $\subs{\overline{\Delta
        s}}{bond}$, as a function of system size.
    }
}
\end{centering}
\end{figure*}

\edit{
The previous section presented complexity measures evaluated on a network
formed from the mutual information $M_{jk}$.  $M_{jk}$ depends on just two
sites.  We expand our study to larger subsystems, finding further evidence
that Goldilocks rules generate complexity.  Large subsystems have von Neumann
entropies that have not been measured experimentally. However, the second-order
R\'enyi entropy is accessible~\cite{islam2015measuring}. This section therefore
focuses on the second-order R\'enyi entropy.
}

\edit{
The entropy is defined as follows.  Consider partitioning the
lattice at its center.  The lattice's halves have equal entropies,
associated with the bond between them and the \emph{bond entropy},
$\subs{s}{bond}$~\cite{Li_2019_measurement}.  If one lattice half occupies
the quantum state $\hat{\rho}$, $\subs{s}{bond} := - \log_2 \big( {\rm Tr}
( \hat{\rho}^2 ) \big)$.  To bridge this measure to complexity science,
we consider fluctuations in $\subs{s}{bond}$.  Entropy fluctuations
quantify complexity in biological systems~\cite{costa2005multiscale}.
In quantum statistical mechanics~\cite{jaksic2011entropic}, entropy
fluctuations diverge at certain transitions between ordered and
disordered phases~\cite{yurishchev2010entanglement}.  Complexity
sits at the boundary between order and disorder, as explained in
Sec.~\ref{sec:results_network-complexity}.  For these reasons, we interpret
high entropy fluctuations generated by QCA as signaling physical complexity.
}

\edit{
Figure~\ref{fig4}(a) shows $\subs{s}{bond}$ as a function of time.  The curves
generated by the Goldilocks rules $T_6$ and $F_4$ (the ruby and amber
curves) fluctuate the most and persistently.  Figure~\ref{fig4}(b) shows
the fluctuations $\Delta \subs{s}{bond}$ plotted against time, for a 19-site
system.  The fluctuations produced by the Goldilocks rules remain at least an
order of magnitude above the fluctuations produced by the non-Goldilocks rules.
Entropy fluctuations therefore signal Goldilocks' rules complexity.
}

\edit{
This behavior remains robust as the system grows.  Let $\overline{ \Delta
\subs{s}{bond} }$ denote the entropy fluctuations averaged over late times.
Figure~\ref{fig4}(d) shows $\overline{ \Delta \subs{s}{bond} }$ plotted against
the system size.  As $L$ grows, the entropy fluctuations shrink most quickly
for the random state (black dots) and the far-from-Goldilocks rule $F_{26}$
(green dots), less quickly for the near-Goldilocks rules (yellow and blue
dots), and most slowly for the Goldilocks rules (ruby and amber dots).
Hence Goldilocks rules generate complexity robustly, according to entropy
fluctuations.
}

\edit{
As an aside, although the Goldilocks rules produce entanglement, according to
Figs.~\ref{fig4}(a) and~\ref{fig4}(c), they produce the least.  This behavior
might remind one of many-body scars---nonequilibrating features discovered
under continuous-time $T_1$ evolution.  Many-body scars mark a few area-law
energy eigenstates states amidst a volume-law sea. Yet observing many-body
scars requires special initial conditions.  Fine-tuning is not necessary for
Goldilocks rules to hinder the entropy's equilibration: Complexity emerges
for diverse initializations, activation operators, and boundary conditions, as
shown in App.~\ref{sec:supplemental}.
}

\subsection{\label{sec:results_breather}
    Entangled breather}

\begin{figure*}[t]
\begin{centering}
\includegraphics[width=1.0\textwidth]{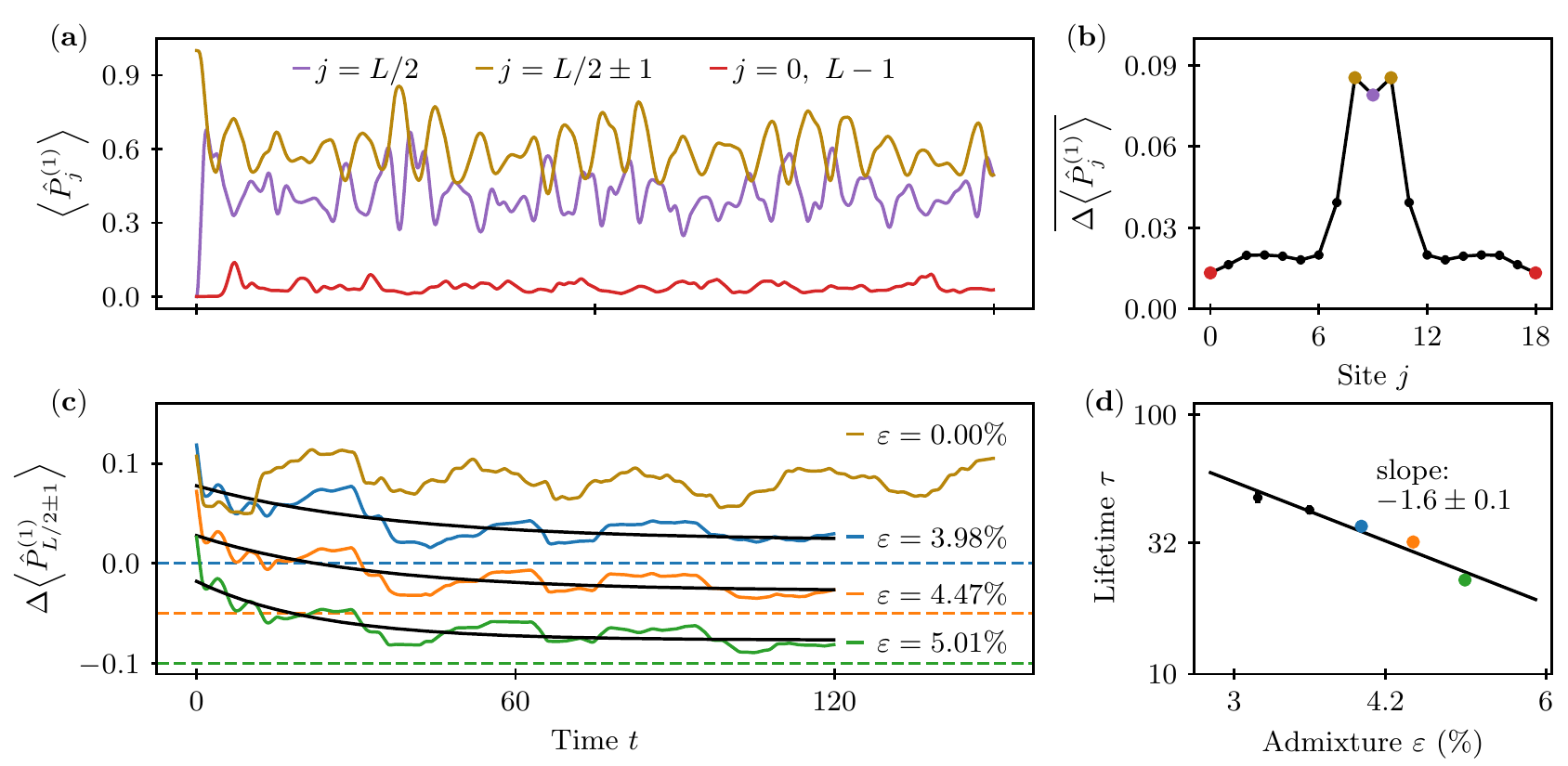}
\caption{\label{fig5}
    \edit{
    The entangled breather produced by Goldilocks rule $F_4$ has a
    characteristic structure and is robust to perturbations. Here we analyze
    the breather on an $L{=}19$-site lattice and denote the center site by
    $L/2$.
    (a) The probability of site $j$ being in $\ket{1}$,
        $\braket{\op{P}^{(1)}_j}$, shows temporal oscillations occur with the
        greatest amplitude near the lattice's center.
    (b) Late-time average of fluctuations in the probability of sites being in
        $\ket{1}$, $\overline{\Delta \braket{\op{P}_j^{(1)}}}$, quantifies the
        spatial localization of temporal oscillations. The coloring is as in
        panel (a).
    (c) Upon perturbation by the $F_{26}$ Hamiltonian, scaled by $\varepsilon$,
        the characteristic fluctuations near the lattice's center decay
        exponentially. The black lines show exponential fits that yield a
        lifetime $\tau$. The decaying curves are shifted by 0.05 from one
        another to aid visualization. The dashed horizontal lines mark the
        shifted $x$-axes.
    (d) The entangled breather lifetime decays as as a function of the
        Hamiltonian-perturbation strength according to a power law. The
        coloring is as in panel (c).
    }
}
\end{centering}
\end{figure*}

\edit{
The sixth signal of Goldilocks rules' complexity is a robust
emergent phenomenon.  The emergent phenomena of life and
consciousness signal complexity most famously.  Other examples
appear in economic markets~\cite{arthur1999complexity},
Internet traffic~\cite{takayasu2000dynamic},
physiological airways~\cite{suki2011complexity}, and neural
activity~\cite{chialvo2010emergent}.  Figures~\ref{fig2}(c)-(d) display
the entangled breather produced when the Goldilocks rule $F_4$ evolves the
initial state $\ket{0 0 \dots 0 \, 101 \, 00 \dots 0}$.  Here, we analyze the
breather's dynamics and robustness.
}

\edit{
Breathers fall under the purview of discrete nonlinear wave theory. There,
a \emph{breather} is defined as a solution that oscillates in time and
that is localized spatially~\cite{flach2008discrete, lederer2008discrete}.
Let us establish that the phenomenon we observe satisfy these criteria.
Imagine measuring the $\op{\sigma}^z$ of each site $j$.  The probability of
obtaining the +1 outcome is $\braket{\op{P}_j^{(1)}}$, wherein $\op{P}_j^{(1)}
:= \ket{1}_{j}\bra{1}_j$.  Figure~\ref{fig5}(a) shows $\braket{\op{P}_j^{(1)}}$
plotted against time for key sites $j$.  The central site is at $j=\lfloor L/2
\rfloor$, whose notation we simplify to $L/2$.  The central site ($j = L/2$,
purple curve) and its two nearest neighbors ($j = L/2 \pm 1$, amber curve)
exhibit $\braket{\op{P}_j^{(1)}}$'s that oscillate throughout the evolution.
We have confirmed the first defining property of breathers.
}

\edit{
To confirm the second property, we compare the oscillations near the
lattice's center to the oscillations at the edges ($j = 0, L-1$, red curve).
The central oscillations have much greater amplitudes.  We quantify this
comparison with the fluctuations $\Delta \braket{\op{P}_j^{(1)}}$, the
temporal-standard deviation across a sliding window of $L$ time units.
Averaging the fluctuations over late times produces $\overline{\Delta
\braket{\op{P}_j^{(1)}}}$, plotted against $j$ in Fig.~\ref{fig5}(b).
The fluctuations peak strongly near the lattice's center.  Hence the emergent
phenomenon is strongly localized and so is a breather.
}

\edit{
Two sets of evidence imply the breather's entanglement.  One consists
of the entanglement measure in Section~\ref{sec:results_bond-entropy}.
(Section~\ref{sec:results_network-complexity} displays results consistent with
entanglement, although the mutual information captures classical correlations
in addition to entanglement.)  Second, the Schmidt-truncation study described
later in this section implies the breather's entanglement.
}

\edit{
This emergent phenomenon could be delicate; its lifetime could be exponentially
suppressed by errors in, e.g., an experiment. The complexity generated by
the Goldilocks rule $F_4$ would be tenuous. We demonstrate, however, that
the entangled breather is robust with respect to perturbations of three
types: perturbations of the Hamiltonian, of the entanglement, and of the
initalization.  Left unperturbed, the entangled breather oscillates for as long
as our numerics run.  To quantify the breather's robustness, we perturb the
breather and measure the center site's neighbors ($j = L/2 \pm 1$).  We fit the
fluctuations $\Delta \braket{\op{P}_{L/2 \pm 1}^{(1)}}$ with the exponential
$A+Be^{-t/\tau}$ to extract breather's lifetime, $\tau$.
}

\edit{
The first perturbation is of the Hamiltonian.  We scale the far-from-Goldilocks
$F_{26}$ Hamiltonian with a positive positive $\varepsilon \ll 1$ and add
the product to the $F_4$ Hamiltonian.  The perturbed dynamics produce an
entangled breather whose oscillations decay.  Figure~\ref{fig5}(c) shows the
fluctuations $\Delta \braket{\op{P}_{L/2 \pm 1}^{(1)}}$ and their fits for
various $\varepsilon$.  The lifetime $\tau$ decays with $\varepsilon$ as the
power law $\tau \sim \varepsilon ^{-1.6}$, as shown in Fig.~\ref{fig5}(d).
The power law decays more slowly than an exponential, indicating the entangled
breather's robustness with respect to Hamiltonian perturbations.
}

\edit{
Second, we perturb the QCA via Schmidt truncation: We evolve the system with a
matrix-product-state approximation.  We reduce the number of singular values
used to represent the state, $\chi$, from $2^{\lfloor L/2 \rfloor} = 2^{8}$
until the breather becomes unstable.  This truncation time-adaptively caps
the amount of entanglement in the state~\cite{jaschke2018open}.  Under Schmidt
truncation, the lifetime $\tau$ changes sharply at $\chi \sim 11$. Below the
threshold, $\tau \propto e^{0.6\chi}$.  At the threshold, $\tau \sim 340$.
Above, $\tau$ exceeds the simulation time of $6,000$ times units. Hence we 
regard the entangled breather as robust with respect to Schmidt truncation.
}

\edit{
Third, we initialized the lattice imperfectly: The $\ket{1}$'s were replaced
with copies of $e^{i\phi} \, \delta \ket{0} + \sqrt{1-\delta^2} \ket{1}$,
wherein $\phi \in [0, \, 2 \pi)$ and $0<\delta \leq 1$.  Whenever $\delta
\leq 1/2$, the breather's lifetime exceeds the simulation time, regardless
of $\phi$. When $\delta > 1/2$, the breather's oscillations still persist
indefinitely, but with a smaller amplitude.  The entangled breather is
therefore robust also with respect to a wide range of perturbations to the
initialization. This emergent phenomenon, with the bond-entropy fluctuations
and the four mutual-information measures, points to the Goldilocks rules'
generation of complexity.
}

\section{\label{sec:models}
    Model details}

In this section, we present formulae for the $T_R$ unitaries and the $F_R$
Hamiltonians, as functions of rule number. Our rule numbering is inspired by
elementary-cellular-automata rule numbering, which we review first.

\subsection{Classical-rule numbering}
A classical-ECA neighborhood consists of a central bit and its two nearest
neighbors.  Classical ECA rules are denoted by $C_R$.  The rule number $R$
encodes the local transition function, which prescribes a bit's next state
for every possible neighborhood configuration.  A bit has two possible values
(0 or 1), and ECA neighborhoods consist of three sites.  Hence a neighborhood
occupies one of $2^3=8$ configurations. Each configuration can lead to the
central bit's being set to 0 or 1; so $2^{2^3}=256$ $C_R$ rules exist.

The system is evolved through one time step as follows.  First, a copy of the
bit string is produced.  Each site's neighborhood is read off the copy and
inputted into the local transition function.  The function's output dictates
the central site's next state, which is written to the original bit string.
The copying allows all sites to be updated independently, simultaneously.

\begin{figure}[t]
    \centering
    \includegraphics[width=0.45\textwidth]{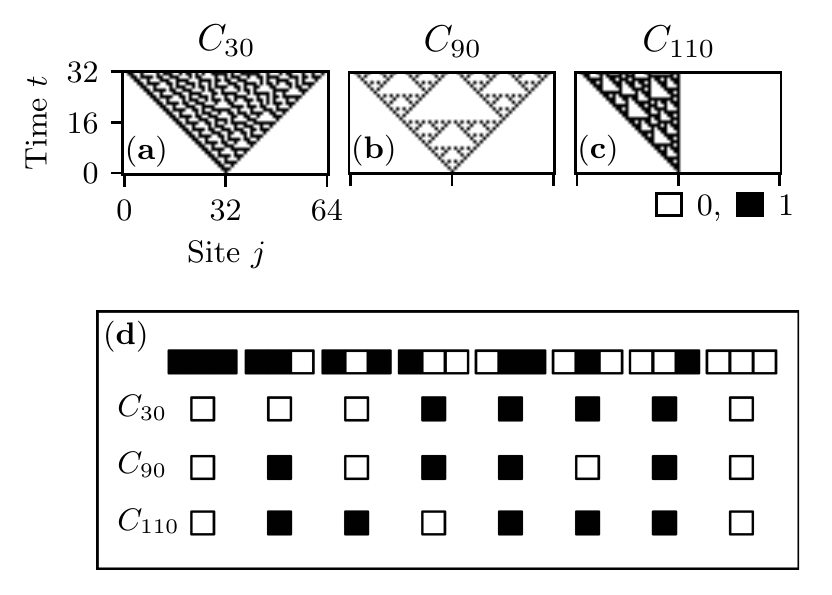}
 \caption{\label{fig:classical_ECA}
    (a)-(c) Example classical-elementary-cellular-automata (ECA) evolutions
          of a 1 centered in 0s.
    (d) Rule tables for $C_{30}$, which generates pseudorandom
        numbers~\cite{wolfram1985cryptography}; $C_{90}$, which generates a
        fractal structure; and $C_{110}$, which is Turing-complete (capable of
        simulating any computer program)~\cite{cook2004universality}.
    }
 \end{figure}

The label $C_R$ encodes the rule as follows.  Given $C_R$, expand $R$ into
8 binary digits, including leading zeroes: $R = \sum_{n=0}^7 c_{n} 2^n$.
The index $n$ is a base-10 number that can be translated into binary: $n =
\sum_{m=0}^2 k_m 2^k = k_2 k_1 k_0$.  These three bits represent a possible
initial configuration of the neighborhood.  If the neighborhood begins in the
$n^{\rm th}$ possible configuration, the central site is updated to the $c_n
\in \{0,1\}$ in the binary expansion of $R$.  Figure~\ref{fig:classical_ECA}
shows the update table for rules $R = 30$, 60, and 110.  Also depicted are the
time evolutions of a 1 centered in 0s.

\subsection{\label{sec:Def_T}
    Digital quantum cellular automata}

$T_R$ denotes a digital 3-site-neighborhood QCA rule.  To expose the
rule's dynamics, convert $R$ from base-10 into four base-2 bits (including
leading zeroes): $R = c_{11} 2^3 + c_{10} 2^2 + c_{01} 2^1 + c_{00} 2^0 =
\sum_{m,n=0}^1 c_{mn}2^{2m+n}$.  The $m$ specifies the state of a central
site's left-hand neighbor, and $n$ specifies the right-hand neighbor's state.
$c_{mn} \in \{0, 1\}$ specifies whether the central bit evolves under
a single-qubit unitary $\op{V}$ (when $c_{mn}=1$) or under the identity
$\op{\Id}$ (when $c_{mn}=0$), given the neighbors' states.  The three-site
update unitary has the form
\begin{equation} \label{eq:3siteU}
    \op{U}_j(\op{V}) = \sum_{m,n=0}^1
        \op{P}^{(m)}_{j-1} \,
        \op{V}^{c_{mn}}_j \,
        \op{P}^{(n)}_{j+1} \, .
\end{equation}
The superscript on $\op{V}_j$ denotes a matrix power.  We update the lattice
by applying $\op{U}_j(V)$ to each 3-site neighborhood centered on site $j$,
for every $j$.  Even-$j$ neighborhoods update first; and odd-$j$ neighborhoods,
second.  One even-site update and one odd-site update, together, form a time
step.

\subsection{Analog quantum cellular automata}

$F_R$ denotes a totalistic analog QCA rule that updates each site $j$,
conditionally on four neighbors, with a Hermitian activation operator
$\op{h}_j$. The value of $R$ dictates the form of the many-body Hamiltonian
$\op{H} = \sum_j \op{H}_j$.  We expand $R$ into 5 binary digits: $R =
\sum_{q=0}^4 c_q 2^q$.  Site $j$ evolves conditionally on its neighbors through
\begin{equation} \label{eq:5siteH}
    \op{H}_j = \op{\sigma}^x_j \sum_{q=0}^{4} c_q
                    \op{\mathcal{N}}_j^{(q)} \, .
\end{equation}
$\op{\mathcal{N}}_j^{(q)}$ denotes the projector onto the $q$-totalistic
subspace, in which exactly $q$ neighbors are in $\ket{1}$'s:
\begin{equation} \label{eq:qprojector}
    \op{\mathcal{N}}_j^{(q)} =
    \sum_{ \substack{ \alpha,\beta,\gamma,\delta \in \{0,1\}, \\
                      \alpha+\beta+\gamma+\delta = q} }
        \op{P}_{j-2}^{(\alpha)} \op{P}_{j-1}^{(\beta)}
        \op{P}_{j+1}^{(\gamma)}\op{P}_{j+2}^{(\delta)} \, .
\end{equation}
The sum runs over $\binom{4}{q}$ permutations of the neighboring $\ket{0}$'s
and $\ket{1}$'s.  We report on $\op{h}_j = \op{\sigma}_j^x$ for reasons
explained in Sec.~\ref{sec:Scope}.

\section{ \label{sec:physical}
    Physical implementation of quantum cellular automata}
Having established the QCA's rule numbering, we prescribe a scheme for
implementing QCA physically.  We focus on analog QCA because they naturally
suit our target platform, an analog quantum simulator.  In this section, $T_R$
refers to an analog 3-site QCA.  Appendix~\ref{sec:analog_digital_relationship}
details the relation between analog and digital $T_R$ protocols.

We begin with general results, before narrowing to an example platform.  First,
we show that the longitudinal-transverse quantum Ising model can reproduce
QCA dynamics.  Many experimental platforms can simulate this Ising model and
so can, in principle, simulate QCA.  Examples include superconducting qubits,
trapped ions, and Rydberg atoms.

We illustrate our experimental scheme with neutral atoms trapped in an
optical-tweezer lattice and excited to Rydberg states with principle
quantum numbers $\gg 1$.  Excited atoms interact through a van der
Waals coupling that induces a Rydberg blockade: Two close-together atoms
cannot be excited simultaneously.  We exploit this blockade to condition
a site's evolution on the site's neighbors.  Moreover, Rydberg atoms
can be arranged in various geometries subject to short- and long-range
interactions~\cite{adams2019rydberg}.  We exploit these freedoms to engineer
nearest-neighbor rules with a linear atom chain and next-nearest-neighbor rules
with a ladder-like lattice [Fig.~\ref{fig:suppl_ryd}(k)].

Related work appeared in the literature recently: A Rydberg-chain realization
of QCA was proposed in~\cite{wintermantel2020unitary}
\footnote{
    Operations reminiscent of QCA were also implemented
    in~\cite{fitzsimons2007quantum}, albeit without mention of QCA.
}.
Wintermantel \emph{et al.} focus on 3-site neighborhoods. We present a
more-general version of unitary Rydberg-atom QCA, which encompasses 3-site and
5-site neighborhoods.

This section is organized as follows.  In Sec.~\ref{sec:QCA-Ising}, we map
analog QCA onto the Ising Hamiltonian. Section~\ref{sec:Ising-Ryd} shows how to
implement this Hamiltonian with Rydberg atoms.  We simulated these Rydberg-atom
QCA numerically, obtaining results presented in Sections~\ref{sec:Ryd-T}
and~\ref{sec:Ryd-F}.

\subsection{\label{sec:QCA-Ising}
    From QCA to the Ising model}

Analog QCA involve the Hermitian activation operator $\op{h}_j$, which
we have chosen to be $\op{\sigma}_j^x$ in this paper.  According to
Eq.~\eqref{eq:QCAana}, $\op{h}_j$ evolves site $j$ if the neighboring sites
satisfy the conditions imposed by the projectors.  Such a state change can be
modeled with the Ising Hamiltonian\edit{: The projectors are implemented with
local energy penalties that shift undesired transitions far off-resonance.}

Let $h_x$ denote the transverse field; and $h_z$, the longitudinal field.
The Ising Hamiltonian has the form
\begin{equation}\label{eq:ham_ising}
    \op{H} = h_x \sum_j \op {\sigma}_j^x
    + h_z \sum_j \op{\sigma}_j^z
    + \frac{J}{2}\sum_j\sum_{k\in \Omega_j} \op{\sigma}_j^z
    \op{\sigma}_{k}^z\, .
\end{equation}
\edit{
The interaction strength $J$ is uniform across the neighbors $k \in \Omega_j$
and vanishes outside.  If $h_x=0$, the resulting Hamiltonian, $\hat H_0$, can
be represented by a block-diagonal matrix relative to the $\hat{\sigma}^z$
product basis.  Each block corresponds to one total-energy value.  Consider
switching on $h_x\ll J \approx h_z$. Some transitions between $\hat H_0$
eigenstates are resonant in energy, while others require energy jumps $\sim J$,
depending on the relationship between $J$ and $h_z$.  An initial state in one
$\hat H_0$ block will evolve within the block.  By tuning $h_z$ and $J$, we can
impose the resonance condition for a set transitions.
}

\edit{
Now, we show how to choose values for $h_z$ and $J$.  Spin $j$ evolves under
the influence of the other spins, in an interacting global quantum system.
However, we approximate $j$ as evolving in a classical field.  $h_x$ serves
as the $j^{\rm th}$ qubit's intrinsic gap.  $h_z+J m$ serves as the effective
external field's frequency. \edit{The effective external field creates an
energy penalty dependent on the}  magnetization, $m = \braket{\sum_{k\in
\Omega_j} \op{\sigma}_k^z}$.  The magnetization counts the $\ket{0}$'s in
$\Omega_j$ and subtracts off the number of $\ket{1}$'s.
}

Under each of several $T_R$ rules, the activation operates only if a
certain number of neighbors occupy $\ket{1}$'s: $m$ must have one particular
value. Examples include $T_1$, $T_6$, and $T_8$.  The Ising model can implement
these rules if the longitudinal field is $h_z = -m J$.  The central spin's
oscillations will be resonant with the effective external field at the
required $m$ value.  If $m$ does not have the required value, the central
spin's oscillations are strongly off-resonant.  This off-resonance effects the
projectors in Eq.~\eqref{eq:QCAana}.  The central spin's maximum probability
of flipping, across a Rabi cycle, is  $h_x/\sqrt{h_x^2+(h_z + J m)^2}$.
Rules that activate at multiple $m$ values would require additional fields
and resonances~\cite{wintermantel2020unitary}. Examples include rule $T_{14}$,
which activates when $m=0, 2$.

\subsection{\label{sec:Ising-Ryd}
    From the Ising Model to the Rydberg-atom quantum simulator}

\edit{
Rydberg atoms are natural platforms for experimental implementations of
the Ising model \cite{Schauss2018c} in the parameter regime described
above. Indeed, the Rydberg blockade naturally realizes the energy penalty
necessary for implementing a QCA. The blockade has been experimentally
observed \cite{bernien2017probing} and recently exploited to simulate
lattice gauge theories \cite{PhysRevX.10.021041, PhysRevX.10.021057,
PhysRevResearch.2.013288}.
}
Each atom's ground state, $\ket{g}$, is coupled to a highly excited Rydberg
state, $\ket{r}$, \edit{via the Rabi frequency $\Omega$}.  Atoms $j$ and
$k \neq j$ experience the isotropic van der Waals interaction $V_{jk}= C_6
/a_{jk}^6$.  The $C_6>0$ denotes an experimental parameter, and $a_{jk}>0$
denotes the interatomic separation.  We use the notation $V_a = V_{jk}$ for
all $\lvert j-k \rvert = 1$ and $\subs{V}{nn} = V_{jk}$ for all $\lvert j -
k \rvert = 2$.

In the next two sections' numerical simulations, we include the
next-highest-order terms, $\subs{V}{ho}$, in the van der Waals interaction
[dashed lines in Figs.~\ref{fig:suppl_ryd})(j)-(k)].  These $\subs{V}{ho}$, we
prove, do not affect the results significantly.

Rydberg atoms evolve under the effective Hamiltonian \cite{bernien2017probing}
\begin{equation} \label{eq:ham_ryd}
    \op{H}_{\rm Ryd.} =
        \frac{\Omega}{2}\sum_j \op{\sigma}_j^x
        + \Delta\sum_j \op{n}_j
        + \frac{1}{2}\sum_{j,k}V_{j,k} \op{n}_j \op{n}_k\, .
\end{equation}
$\Delta$ denotes a detuning in the natural Hamiltonian that is transformed and
approximated to yield Eq.~\eqref{eq:ham_ryd}.  The detuning affects nearly all
the atoms uniformly: The QCA, recall, has boundary sites set to $\ket{0}$'s.
We implement these boundary conditions by tailoring $\Delta$ for the boundary
sites and keeping $\Delta$ constant elsewhere.

\edit{
We map the Rydberg Hamiltonian onto the Ising Hamiltonian as follows. First,
the lattice geometry is chosen such that each site $j$ has $2r$ nearest
neighbors $k \in \Omega_j$.  Second, we set $\Omega = 2 h_x$ and $\Delta =
2(h_z - 2 r J)$. Third, we set the interatomic distance $a$ such that $ V_{a} =
4J$ for the neighbors $k \in \Omega_j$.
}

\edit{
In the Ising Hamiltonian, $V_{j,k} = 0$ for all every site $k$
outside $\Omega_j$. This condition is well-approximated by the Rydberg
Hamiltonian~\eqref{eq:ham_ryd}, due to the van der Waals interaction's
rapid decay with separation.  We fix parameters to reflect the physics of
$^{87}\mathrm{Rb}$ atoms with the excited state $\ket{r}=\ket{70 S_{1/2}}$, for
which $C_6=863~\rm GHz~\mu m^6$.  The interatomic distance $a\simeq 5.4~\rm \mu
m$, and $\Omega=2~\rm MHz  \gg  V_a=36~\rm MHz$~\cite{bernien2017probing}.
}

\subsection{\label{sec:Ryd-T}
    Simulating $T_R$ rules with the Rydberg Hamiltonian}

\edit{
Here, we specify the parameters needed to simulate rules $T_6$, $T_1$ and $T_8$
with the Rydberg Hamiltonian.  Consider a linear atom chain, with interatomic
distance $a$, in the Rydberg-blockade regime [Fig.~\ref{fig:suppl_ryd}(j)].
Appropriately setting the detuning $\Delta$ can enhance the oscillations
undergone by a neighborhood's central spin, conditionally on the spin's
neighbors.
}

The analog version of rule $T_1$ imposes the Hamiltonian
\begin{equation}\label{eq:ham_T1}
    \op{H} = \sum_j \op{P}^{(0)}_{j-1}
    \op{\sigma}_j^x \op{P}^{(0)}_{j+1} \, .
\end{equation}
The Ising Hamiltonian implements the projectors when the resonance condition
is imposed between $\ket{000}_{\Omega_j}$ and $\ket{010}_{\Omega_j}$.  In each
of these states, two neighbors occupy $\ket{0}$'s; two neighbors point upward,
so the magnetization is $m=2$.  Hence the resonance condition implies that
$h_z=-2 J$, or $\Delta=-2V_a$ for the Rydberg Hamiltonian.  Rule $T_8$ effects
the Hamiltonian~\eqref{eq:ham_T1}, except each $\op{P}^{(0)}$ is replaced with
a $\op{P}^{(1)}$.  Hence $m=-2$, and $h_z=2J$, i.e., $\Delta=0$.

The Goldilocks rule $T_6$ implements the Hamiltonian
\begin{equation}\label{eq:ham_T6}
\op{H} = \sum_j \left ( \op{P}^{(0)}_{j-1}
                        \op{\sigma}_j^x
                        \op{P}^{(1)}_{j+1}
                      + \op{P}^{(1)}_{j-1}
                        \op{\sigma}_j^x
                        \op{P}^{(0)}_{j+1} \right ) \, .
\end{equation}
The resonance must be enhanced when the neighboring sites satisfy $m=0$.  Hence
we set $h_z=0$ or, equivalently, $\Delta=-V_a$ in the Rydberg Hamiltonian.

\edit{
Numerical simulations comparing analog QCA and their Rydberg-atom
implementation for $T_1$ are shown in Fig.~\ref{fig:suppl_ryd}(a)-(c).
Figure~\ref{fig:suppl_ryd}(d)-(f) shows the comparison for $T_6$.  The QCA
simulations (in each figure triplet's leftmost column) agree qualitatively with
the Rydberg simulations (depicted in the center column).  Each figure triplet's
rightmost column compares the QCA and Rydberg simulations on the same axes
at key sites.  The agreement exists despite the small, high-order interaction
terms $\subs{V}{ho}\sim V_a/64$ [Fig.~\ref{fig:suppl_ryd}(j)].
}

\subsection{\label{sec:Ryd-F}
    Simulating $F_R$ rules with the Rydberg Hamiltonian}
 \begin{figure*}[ht]
    \centering
    \includegraphics[width=1.0\textwidth]{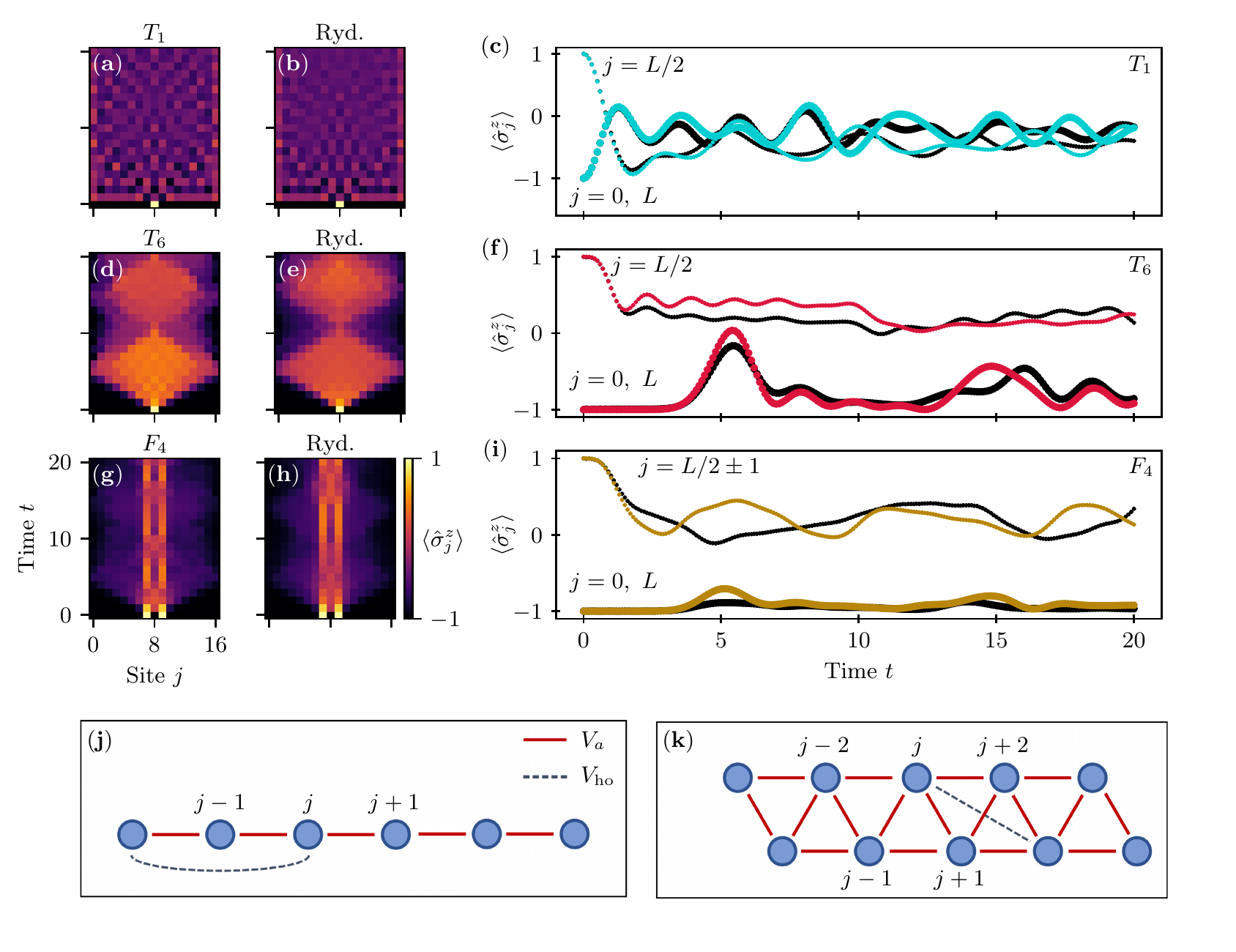}
 \caption{ \label{fig:suppl_ryd}
    \edit{
    (a)-(i) Comparison of QCA evolution and Rydberg-atom dynamics.
            We study $\braket{\op{\sigma}^z_j}$ for the near-Goldilocks
            three-site rule $T_1$ in (a)-(c), the three-site Goldilocks
            rule $T_6$ in (d)-(f) and the five-site Goldilocks rule $F_4$ in
            (g)-(i). Panels (c), (f), and (i) show the results for key sites
            $j$, distinguished by line style. Colored lines reflect the exact
            QCA results; black lines represent their Rydberg-atom simulations.
    (j)-(k) Lattice geometries for the three-site $T_R$ rules (j) and
            five-site $F_R$ rules (k). Blue circles represent the atoms;
            red lines, the interactions with strength $V_a$; and gray, dashed
            lines, the next-nearest-neighbor interactions (under $T_R$ rules)
            and next-next-nearest neighbor interactions (under $F_R$ rules),
            which generate higher-order effects.
    }
 A local detuning is applied at the boundaries to simulate the QCA
 boundary conditions.  Parameters: $L{=}17$.  In the Rydberg simulations,
 $\Omega=2~\mathrm{MHz}$, and $V_a=36~\mathrm{MHz}$.  Time is expressed in
 units of $2 \pi \; \mu \mathrm{s}$.
 }
 \end{figure*}

Simulating the $F_R$ rules with Rydberg atoms requires that excited
next-nearest-neighbor atoms have the same interaction energy as
nearest-neighbor atoms.  One can meet this condition by forming a zigzag
lattice, shown in Fig.~\ref{fig:suppl_ryd}(k).  In this arrangement, the
distance $a_{jk}$ for all $\lvert j-k \rvert = 1$ is equal to the  distance
$a_{jk}$ for all $\lvert j-k \rvert = 2$.  Hence the interaction energies are
equal: $V_{\rm nn}=V_{a}$.

We discuss, as an example, the implementation of the Goldilocks rule $F_4$.
According to this rule, a neighborhood's central spin changes state if 2 of
its 4 neighbors are in $\ket{1}$'s.  Hence, as for $T_6$, $m=0$, leading to
the resonance condition $\Delta=-2 V_a$.  Figures~\ref{fig:suppl_ryd}(g)-(i)
depict the numerical results, and Fig.~\ref{fig:suppl_ryd}(k) depicts the
atomic arrangement.  The plots confirm strong qualitative agreement between
the QCA dynamics and the quantum-simulator dynamics. \edit{Due to the lattice
geometry, higher-order interaction terms are larger for $F_4$ than for $T_1$
and $T_6$: In the $F_4$ case, $V_{ho}\sim V_a/27$. Consequently, the QCA
dynamics diverge from the Rydberg dynamics more: The entangled breather remains
in the Rydberg simulation, but with about half the oscillation frequency. We
have checked that, if the higher-order interaction terms are suppressed, the
agreement is significantly improved, supporting the mapping between the Rydberg
Hamiltonian~\eqref{eq:ham_ryd} and the QCA Hamiltonian.}

 \edit{
The Rydberg-atom dynamics agree visually with the analog QCA. We
quantify the agreement with the normalized absolute deviation $\lvert
\subs{\braket{\op{\sigma}^z_j}}{QCA}-\subs{\braket{\op{\sigma}^z_j}}{Ryd.}
\rvert/2$, averaged over sites $j$ and times $t\in[0,~20]$.  The results are
3.9\% for $T_1$, 5.9\% for $T_6$, and 4.2\% for $F_4$---all errors of, or
slightly above, order 1\%.
}
Thus, Rydberg atoms offer promise for quantum simulation of QCA, for the
observation of complexity's emergence under Goldilocks rules, and for studying
QCA's robustness with respect to physical-implementation details.

\section{\label{sec:conclusions}
    Conclusions}

We have bridged the fields of QCA and complexity science, discovering the
emergence of physical complexity under Goldilocks rules. Goldilocks rules
balance activity and inactivity. This tradeoff produces structured entangled
quantum states. These states are persistently dynamic and neither uniform nor
random, according to complexity measures.
\edit{
Complexity signatures of Goldilocks rules include high clustering; short
average paths; high disparity fluctuations; broad node-strength distributions;
persistent entropy fluctuations; and robust emergent dynamical features,
entangled breathers.
}

Non-Goldilocks rules still display interesting dynamical features. For example,
the PXP model, a particular case of rule $T_1$, has received much attention. As
a near-Goldilocks rule, $T_1$ produces some complexity. However, the confluence
of many strong complexity signatures distinguishes the Goldilocks rules.

\edit{
This work establishes several opportunities for future research. For example,
whether Goldilocks rules are integrable merits study. This question is
orthogonal to the question of whether Goldilocks rules generate complexity,
the present paper's focus. The present study, however, offers hints.  First,
we searched for nontrivial quantities conserved by the Goldilocks rules. We
found only one: $T_6$ with a Hadamard activation unitary conserves $\sum_j
\op{\sigma}^{z}_j \op{\sigma}^{z}_{j+1}$, whose sum includes the boundaries, be
they fixed or periodic.  Second, we have performed preliminary studies on QCA's
entanglement spectra~\cite{calabrese2008entanglement, chamon2014emergent},
which suggest that Goldilocks rules might lie between integrability and
nonintegrability~\cite{Shaffer_2014_irreversibility}.  Many more measures have
been leveraged to diagnose integrability and chaos~\cite{hashimoto2017out,
cotler2017chaos, khare2020localized, sala2020ergodicity}; these measures merit
scrutiny.
}

\edit{
Furthermore, we observed Goldilocks QCA hinder equilibration
(Sec.~\ref{sec:results}). How does this resistance relate to mechanisms
for avoiding thermalization in quantum systems, e.g., many-body
localization~\cite{alet2018many}, quantum scars~\cite{turner2018weak}, and
Hilbert space fragmentation~\cite{sala2020ergodicity, khemani2020localization}?
Contrariwise, we observed equilibration under non-Goldilocks QCA. Do they obey
the eigenstate thermalization hypothesis~\cite{deutsch1991quantum}?
}

\edit{
Finally, we have focused on finite-size systems accessible to classical
computing, which exist in the physical world and can be simulated numerically.
The thermodynamic limit forms an alternative, idealized setting.  Do the
complexity features observed in our finite systems persist, contract, or grow
richer in the infinite-system-size limit?  Preliminary indications based
on finite-size scaling in Fig.~\ref{fig3}(d) show that at least one of our
complexity measures stays large for Goldilocks rules in the thermodynamic
limit, namely clustering.  Disparity and entropy fluctuations presents a
borderline cases that cannot be resolved on \emph{classical} computers based
on our present studies, as their decay is slow and may in fact be leveling
off to a plateau.  Likewise, given that quantum entangled breathers require
about 6-7 sites including their supporting tails, finite-size scaling based on
multiple breathers as fundamental emergent objects is inaccessible on classical
resources.  For example, to scale to 10 breathers requires 60-70 qubits.  Thus,
understanding complexity in the thermodynamic limit is overall a good case for
usage of quantum computers~\cite{arute2019quantum}, and presents an excellent
topic for future research in NISQ-era computing.
}

\edit{
Aside from these theoretical opportunities, we have demonstrated that extant
digital and analog quantum computers can implement our QCA.  For example,
Rydberg atoms can be arranged to engineer five-site QCA neighborhoods, large
enough to support an entangled breather. % Also, digital quantum computers
can simulate QCA dynamics that support complexity even with three-site
neighborhoods.  Our work therefore uncovers a direction for experimental
quantum computation: to explore the rich features of biological and social
complexity that manifest in simple, abiological quantum systems.
}

\section*{Acknowledgments}

The authors thank Clarisa Benett, Haley Cole, Daniel Jaschke, Eliot Kapit,
Evgeny Mozgunov, and Pedram Roushan for useful conversations.  This work
was performed in part with support by the NSF under grants OAC-1740130,
CCF-1839232, PHY-1806372, and PHY-1748958; and in conjunction with the QSUM
program, which is supported by the Engineering and Physical Sciences Research
Council grant EP/P01058X/1.
This work is partially supported by the Italian PRIN 2017, the Horizon 2020
research and innovation programme under grant agreement No 817482 (PASQuanS).
NYH is grateful for funding from the Institute for Quantum Information and
Matter, an NSF Physics Frontiers Center (NSF Grant PHY-1125565) with support of
the Gordon and Betty Moore Foundation (GBMF-2644), for a Barbara Groce Graduate
Fellowship, and for an NSF grant for the Institute for Theoretical Atomic,
Molecular, and Optical Physics at Harvard University and the Smithsonian
Astrophysical Observatory.

%\textbf{Author contributions}:
% Conceptualization (LDC); Data Curation (LEH, MTJ, DLV, PR); Formal Analysis (LEH, MTJ, DLV, PR, NYH, SN, SM, LDC); Funding Acquisition (SM, LDC); Investigation  (all authors); Methodology (LEH, NYH, NB, SM, LDC); Project Administration (LDC); Resources (LDC); Software (LEH, MTJ, DLV, PR); Supervision (NYH, SM, NB, LDC); Validation (LEH, LDC); Writing - original draft (LEH, LDC); Writing - review \& editing (LEH, MTJ, PR, NYH, NB, SN, SM, LDC).

%\textbf{Competing interests:}
%The authors declare they have no competing interests.
%
%\textbf{Data Availability}:
%All data is available in the manuscript or the supplementary materials.

\clearpage
\appendix

\section{\label{sec:methods}
    Methods and definitions}

Throughout this work, we use various quantifiers of quantum dynamics. This
appendix details those methods and definitions.

\subsection{\label{sec:expval}
    Local expectation values}

Let $\op{\mathcal{O}}_j$ denote an observable of qubit $j$.  The observable has
a dynamical expectation value of $\braket{\op{\mathcal{O}}_j}(t) = \Tr \left
( \op{\rho}_j(t) \op{\mathcal{O}} \right )$, where $\op{\rho}_j(t)$ denotes
the site-$j$ reduced density matrix, calculated by partially tracing over the
whole-system density matrix $\op{\rho}(t) = \ket{\psi(t)}\bra{\psi(t)}$, i.e.,
$\op{\rho}_j(t)=\Tr_{k\neq j}\op{\rho}(t)$.

\subsection{\label{sec:entropy}
    Entropy}
Let $\mathcal{A}$ denote a subsystem in the state $\op{\rho}_\mathcal{A}
=\Tr_{j \notin \mathcal{A}} ( \op{\rho} )$.  The \emph{order-$\alpha$ R\'enyi
entropy} is defined as
\begin{equation} \label{eq:renyi}
    s_\mathcal{A}^{(\alpha)} =
        \frac{1}{1 - \alpha}
            \log_2 \left [ \Tr ( \op{\rho}_{\mathcal{A}}^\alpha ) \right ]
\end{equation}
for $\alpha \in [0, 1) \cup (1, \infty)$.  In the limit as $\alpha \rightarrow
1$, the R\'enyi entropy becomes the \emph{von Neumann entropy}:
\begin{equation} \label{eq:vonneumann}
    s_{\mathcal{A}} \equiv s_\mathcal{A}^{(1)}
        = -\Tr \left ( \op{\rho}_{\mathcal{A}}
           \log_2 \op{\rho}_{\mathcal{A}} \right )
        = -\sum_{i=0}^{\mathrm{dim}(\mathcal{A})} \lambda_i \log_2 (\lambda_i) \, .
\end{equation}
The sum runs over the eigenvalues $\{\lambda_i\}$ of $\op{\rho}_\mathcal{A}$.
The set $\{\lambda_i\}$ of eigenvalues is called the \emph{entanglement
spectrum} of subsystem $\mathcal{A}$.

We distinguish three subsystems for entropy calculations.  First, the
\emph{local entropy} $s_j^{(\alpha)}$ is the $j^{\rm th}$ site's entropy.
Second, the \emph{two-point entropy}, $s_{j,k}^{(\alpha)}$ for $j \neq k$,
is of a pair of sites.  Third, the entropy of the subsystem formed by cutting
the lattice in half is denoted by $\subs{s}{bond}$.  If the system has an
odd number $L$ of sites, the central bond lies between sites $(L-1)/2$ and
$(L+1)/2$.

\subsection{\label{sec:MI}
    Mutual information}

The order-$\alpha$ quantum R\'enyi mutual information between sites $j$ and $k
\neq j$ is
\begin{equation} \label{eq:mutualinfo}
    M_{jk}^{(\alpha)} =
        \frac{1}{2} \left | s_j^{(\alpha)} + s_k^{(\alpha)}
                            - s_{jk}^{(\alpha)} \right | \, .
\end{equation}
The $1/2$ is unconventional but normalizes the mutual information such
that $0\leq M_{jk}^{(\alpha)}\leq1$.  We define $M_{jj}^{(\alpha)} = 0$
and interpret $M_{jk}^{(\alpha)}$ as the adjacency matrix of a weighted,
undirected graph that lacks self-connections.  The absolute value in
Eq.~\eqref{eq:mutualinfo} is necessary only for $\alpha > 1$.

For $\alpha = 1$, the mutual information is positive-definite.
It has the physical interpretation of upper-bounding every two-point
correlator~\cite{wolf2008area}.  Hence a nonzero $M_{jk}^{(1)}$ reflects
two-point correlations. We focus on $M_{jk}^{(1)}$ because of it's
experimental accessibility, owing to the fact that $M_{jk}$ considers
subsystems of two qubits or fewer.  We have found the complexity outcomes
for $\alpha=1, 2$ are qualitatively similar, as in the context of quantum
phase transitions~\cite{Sundar2018Complex}. To simplify notation, we drop
the $(\alpha)$ superscript in Sec.~\ref{sec:results}. An adjacency matrix can
be defined alternatively in terms of two-point correlators, but they capture
complexity less than mutual information does~\cite{Sundar2018Complex}.

\section{\label{sec:analog_digital_relationship}
    Relation between analog and digital time-evolution schemes}

We will prove that evolution under analog three-site QCA is equivalent to
evolution under the digital three-site QCA.  Consider the radius-1 analog
QCA $T_R$, with rule number $R = \sum_{m,n} c_{mn} 2^{2m+n}$.  According to
Eq.~\eqref{eq:QCAana}, the three-site Hamiltonian has the form
\begin{equation} \label{eq:3siteH}
    \op{H}_j = \sum_{m,n=0}^1 c_{mn}
               \op{P}_{j-1}^{(m)} \,
               \op{\sigma}_j^x \,
               \op{P}_{j+1}^{(n)}\, .
\end{equation}
The state vector is evolved for a time $\delta t$ by the propagator $\op{U}
= \exp(- i \delta t \op{H})$.  For all $j$ and $k$ such that $|k-j| \neq 1$,
$[\op{H}_j, \op{H}_k] = 0$.  Hence the propagator factorizes into even and
odd parts at order $\delta t^2$: $\op{U} \approx \left( \Pi_{j\in\{1,3,5,
\ldots \}} \op{U}_j \right) \cdot \left( \Pi_{j\in\{0,2,4, \ldots \}} \op{U}_j
\right)$.  The local propagator has been defined as $\op{U}_j = \exp(-i \delta
t \op{H}_j)$.  This $\op{U}$ is equivalent to the circuit that runs the digital
$T_R$ simulations, up to the unitary activation operator $\op{V}_j$.

Now, we relate the discrete-time scheme's $\op{V}_j$ to the continuous-time
scheme's $\op{h}_j = \op{\sigma}^x$.  We expand the local propagator as
$\op{U}_j = \op{\Id}
            -i \delta t \op{H}_j
            - \delta t^2 \op{H}_j^2/2!
            + i \delta t^3 \op{H}_j^3/3!
            + \ldots$
The projectors are orthogonal, $\op{P}_j^{(m)} \op{P}_j^{(n)} = \delta_{mn}
\op{P}_j^{(m)}$, and the Pauli operators square to the identity:
$(\op{\sigma}^x)^2 = \op{\Id}$.  Therefore, powers of the Hamiltonian can
be expressed as $\op{H}_j^{2k} =\sum_{m,n}c_{mn} \op{P}_{j-1}^{(m)} \op{\Id}
\op{P}_{j+1}^{(n)}$ and $\op{H}_j^{2k+1} = \op{H}_j$ for $k=1,2,\dots$ Hence
the matrix exponential simplifies to
\[
    \begin{aligned}
    \op{U}_j &= \sum_{m,n=0}^1 \op{P}_{j-1}^{(m)}
    \left [ (1-c_{mn}) \op{\Id}+
    c_{mn} \exp ( i \delta t \op{\sigma}^x_j )
    \right ]
    \op{P}_{j+1}^{(n)}\\
    &= \sum_{m,n=0}^1 \op{P}_{j-1}^{(m)}
    \exp ( i \delta t \op{\sigma}^x_j )^{c_{mn}}
    \op{P}_{j+1}^{(n)}.
    \end{aligned}
\]
The last line has the form of the three-site update unitary in
Eq.~\eqref{eq:3siteU}.  The parallel suggests an activation unitary $\exp (
i \delta t \op{\sigma}^x)$.

\section{\label{sec:supplemental}
    Supplemental Material}
\edit{
In this appendix, we provide additional analysis of several digital $T_R$ rules
to support our claim that Goldilocks rules produce physical complexity for a
variety of activation operators, initial states, and boundary conditions.
Additionally, we provide an alternative visualization of the data in
Fig.~\ref{fig2}.  We illustrate the class of non-Goldilocks rules with
$T_{13}$, a rule not covered in the main text, for the sake of exhibiting a
wider variety of QCA.  $T_{13}$ updates a site $j$ if (i) both neighbors are in
$\ket{0}$'s, (ii) both neighbors are in $\ket{1}$'s, or (iii) the right-hand
neighbor is in $\ket{1}$, while the left-hand neighbor is in $\ket{0}$.
This rule is nontotalistic.
}

\edit{
For a more general activation operator, we consider the product
of a Hadamard and a phase gate: $\op{V}(\upsilon)=\subs{\op{U}}{H}
\subs{\op{U}}{phase}(\upsilon)$. In the matrix representation with $\ket{0} =
\sups{(1,0)}{T}$ and $\ket{1} = \sups{(0,1)}{T}$,
\begin{equation}
    \op{V}(\upsilon) \to \frac{1}{\sqrt{2}}
            \begin{pmatrix}
            1 & e^{i \upsilon} \\
            1 & -e^{i \upsilon}
            \end{pmatrix} \, .
\end{equation}
The main text details the $\upsilon=0$ case.
}

\begin{figure}[ht]
\includegraphics[width=0.45\textwidth]{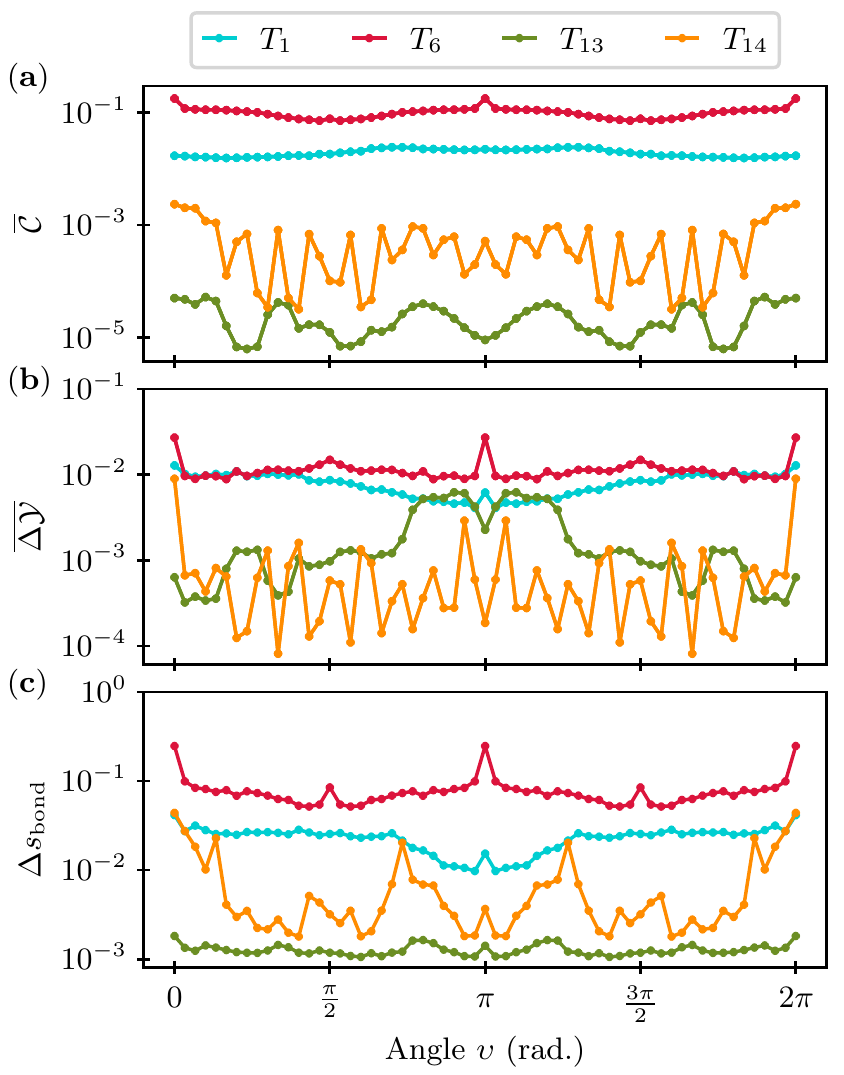}
\caption{\label{fig:phase-gate}
    \edit{
    We generalize the activation unitary from a Hadamard gate to a Hadamard
    gate times an azimuthal phase gate of angle $\upsilon$. The long-time
    trends that distinguish rule $T_6$ remain.  (a) Clustering, (b) disparity
    fluctuations, and (c) bond entropy fluctuations at averaged over times $t \in 500,~1,000$. As in the main text, these simulations are performed with $L=19$
    qubits, an initial condition of a single $\ket{1}$ centered in $\ket{0}$'s,
    and boundary qubits fixed to $\ket{0}$'s.
    }
}
\end{figure}

\edit{
Figure~\ref{fig:phase-gate} shows key complexity metrics: The
mutual-information-network clustering [panel (a)], disparity fluctuations [panel
(b)], and bond-entropy fluctuations [panel (c)] are observed under various rules
and phase-gate angles $\upsilon$. The long-time averages are calculated over the
interval $t\in[500,~1,000]$. The Goldilocks rule $T_6$ exhibits the
largest clustering, disparity fluctuations, and entropy fluctuations for all 
$\upsilon$. Like $T_6$, the near-Goldilocks
rule $T_1$ is, in comparison to the other rules, less sensitive to
nonzero $\upsilon$. Conversely, the near-Goldilocks rule $T_{14}$ and the
far-from-Goldilocks rule $T_{13}$ produce the most variation with $\upsilon$
and the lowest overall values at late times.
}

\begin{figure*}[ht]
\includegraphics[width=1.0\textwidth]{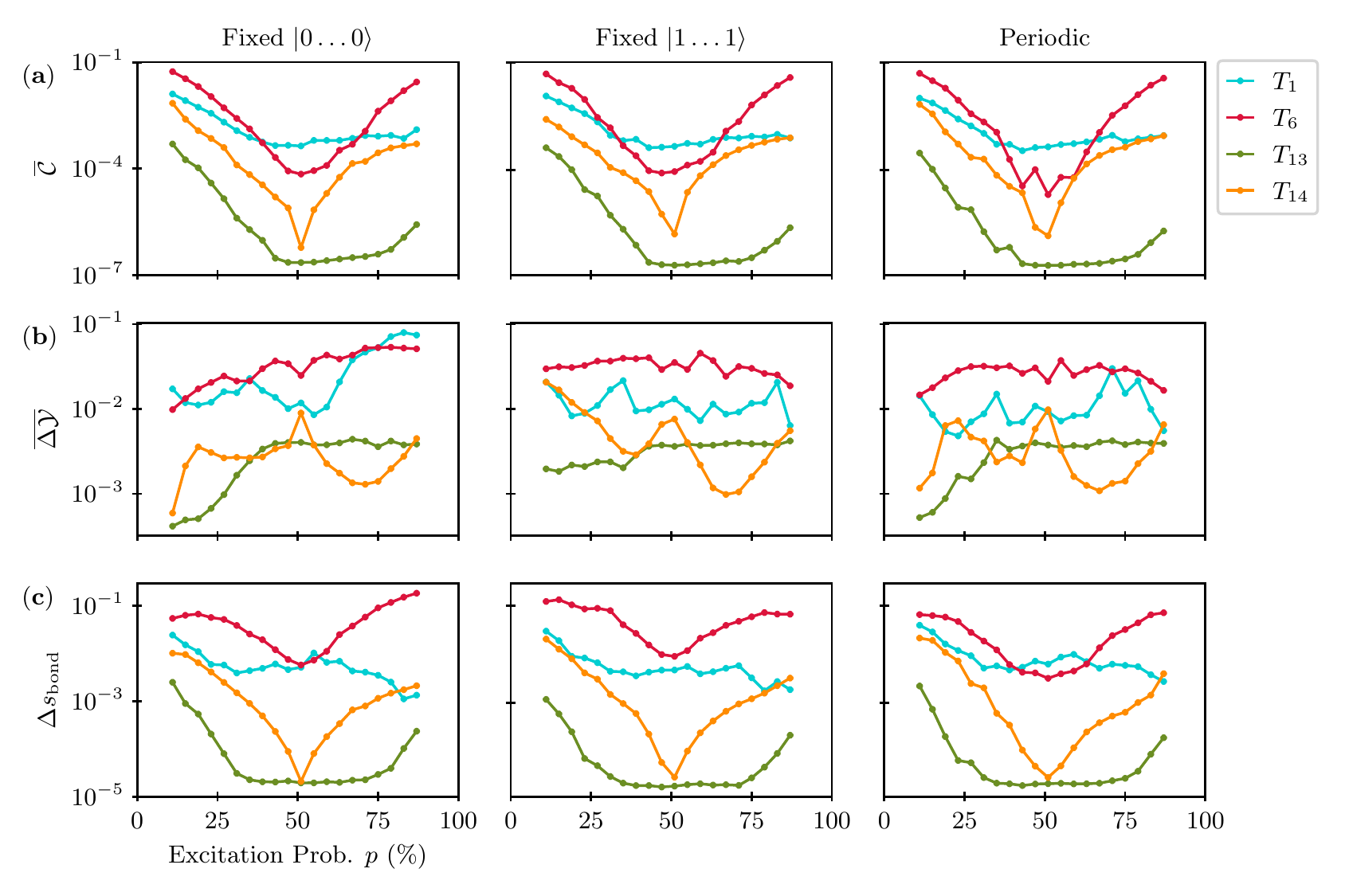}
\caption{\label{fig:IC-BC}
    \edit{
    Goldilocks rule $T_6$ generates the most complexity even under
    various boundary conditions and when initial states are generalized
    to ensembles. The long-time averages of (a) clustering, (b) disparity
    fluctuations, and (c) bond-entropy fluctuations are shown for various
    $T_R$ rules and excitation probabilities $p$. The results are shown
    for boundaries fixed to $\ket{0}$'s (left columns), boundaries fixed to
    $\ket{1}$'s (center column), and periodic boundaries (right column). For
    these simulations, the system size is reduced to $L=15$, and the long-time
    average is taken for $t \in (250, \, 500]$, to mitigate computational
    expense.
    }
}
\end{figure*}

\edit{
Let us next consider the complexity outcomes for alternative initial and
boundary conditions (Fig.~\ref{fig:IC-BC}). We initialize each QCA in a random
product state where each qubit is set to $\ket{1}$ with probability $p$ and
otherwise set to $\ket{0}$. The QCA evolution is performed; and the reduced
density matrices of single sites, two-site pairs, and the central bipartition
are recorded at each time step. The process is repeated 500 times with
different realizations of the initial condition. At each time step, the reduced
density matrices are averaged across the trials. From the ensemble-averaged
reduced density matrices, we compute the long-time-average clustering [row
(a)], disparity fluctuations [row (b)], and entropy fluctuations [row (c)]. We
repeat these simulations for boundaries fixed to $\ket{0}$'s (left column),
boundaries fixed to $\ket{1}$'s (center column), and periodic boundary
conditions (right column).
}

\edit{
Boundary conditions play a relatively insignificant role in determining
complexity outcomes, as compared to initial conditions. Also, the clustering
and and entropy fluctuations are near their minimum, for all rules, with
initial conditions near $p=50\%$. Disparity fluctuations, in comparison,
appear less sensitive to changes in the initial state. The Goldilocks rule
$T_6$ generally maintains the highest clustering, disparity fluctuations,
and entropy fluctuations for all $p$. The exception is near $p=50\%$,
where the near-Goldilocks rule $T_1$ exhibits a greater clustering. The
far-from-Goldilocks rule $T_{13}$ appears the least complex for all initial
conditions. The near-Goldilocks rules $T_1$ and $T_{14}$ lie somewhere between
$T_6$ and $T_{13}$.
}

\begin{figure*}[t]
\begin{centering}
\includegraphics[width=1.0\textwidth]{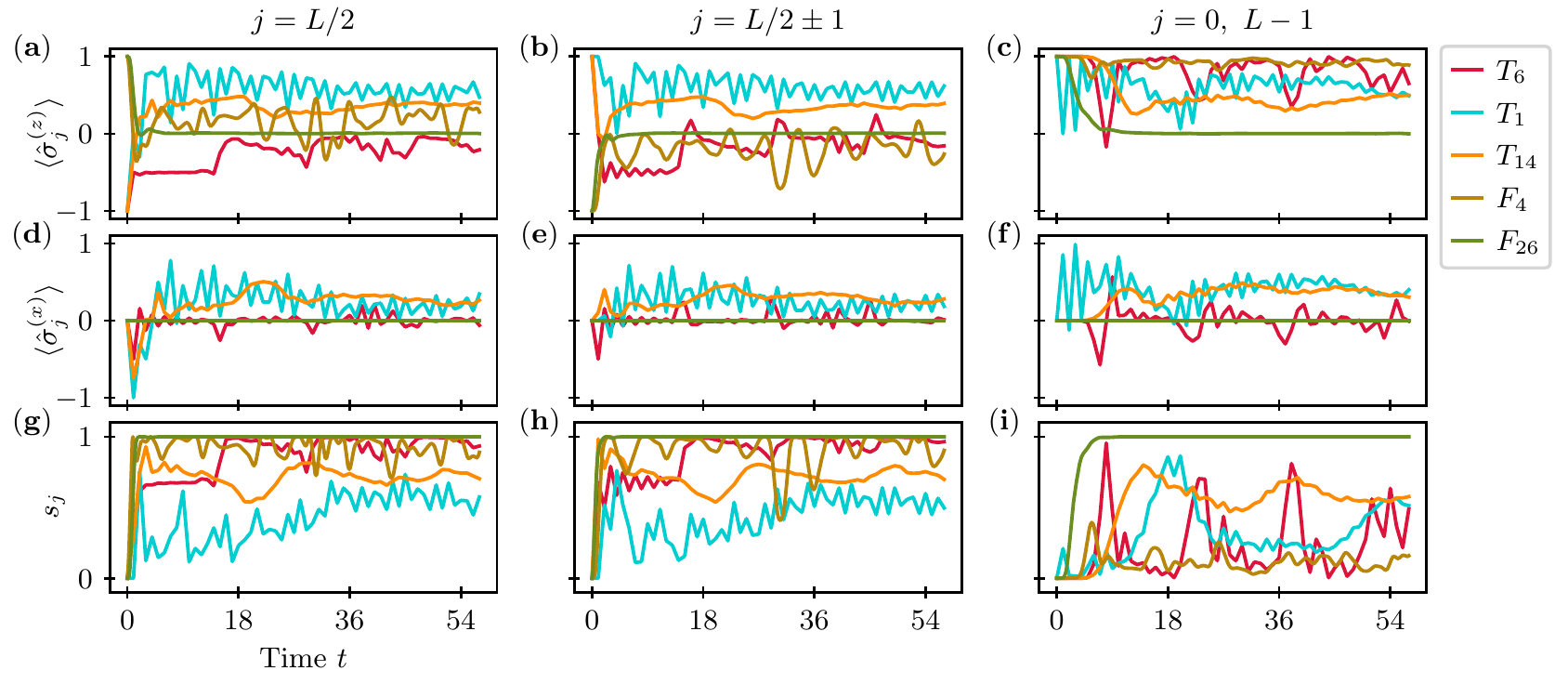}
\caption{\label{fig:linecuts}
    \edit{
    Visualizing the expectation value $\braket{\op{\sigma}_{j}^z}$
    of the Pauli-$z$ operator (a-c), the expectation value
    $\braket{\op{\sigma}_{j}^x}$ of the Pauli-$x$ operator (d-f), and the
    von Neumann entropy $s_{j}$ (g-i).  Each column is labeled by the sites
    $j$ on which the measures are evaluated. The line colors denote the five
    rules considered in the main text.  These simulations are performed for
    $L=19$ qubits and up to 57 time units. The boundary qubits are fixed to
    $\ket{0}$'s. The digital $T_R$ rules use a Hadamard unitary activation
    operator and are initialized to  $\ket{\ldots 000 \: 1 \: 000\ldots}$. The
    analog $F_R$ rules use a Pauli-$x$ Hermitian activation operator and are
    initialized to  $\ket{\ldots 000 \: 101 \: 000\ldots}$.}
}
\end{centering}
\end{figure*}

\edit{
Finally, Figure~\ref{fig:linecuts} shows the one-point dynamics as a
function of time for key sites. This visualization provides an alternative
to Fig.~\ref{fig2} that allows one to compare more directly the site-local
observables for various QCA rules.
}

\clearpage

\bibliography{qca}

\end{document}